\documentclass[12pt,onecolumn,aps,superscriptaddress,
               amsmath,amssymb,nofootinbib,showpacs]{revtex4}      
\usepackage{graphicx}

\usepackage{graphicx,epsfig}

\usepackage{dcolumn}
\usepackage{bm}

\newcommand{\Z}{{\mathbb Z}}

\begin{document}


\title{Light Higgsino Dark Matter in the MSSM on D-branes}

\author{Van E. Mayes}
\affiliation{Department of Physics, University of Houston-Clear Lake, Houston, TX 77058}
\author{Andrew W. Lutz}
\affiliation{Department of Mathematics, The University of Oklahoma,
Norman, OK 73019}

\begin{abstract}
When supersymmetry breaking is dominated by 
the complex structure moduli and the universal dilaton, 
a subset of the supersymmetry parameter space in a realistic MSSM constructed from 
intersecting/magnetized D-branes are universal, similar to the effective 
mSUGRA/CMSSM parameter space with a universal scalar mass $m_0$, a universal gaugino mass
$m_{1/2}$ and with the universal trilinear term fixed to be 
minus the gaugino mass, $A_0=-m_{1/2}$. More generally, the scalar mass-squared
terms for sfermions are split about the Higgs mass-squared terms,
$m_{Q_L,L_L}^2=m_H^2 - \Delta m^2$ and $m_{Q_R,L_R}^2=m_H^2 + \Delta m^2$,
for generic values of the K\a"ahler moduli.  
The scalar masses are universal only for a specific choice of the K\a"ahler moduli.
The hyberbolic branch/focus point (HB/FP)
regions of this parameter space are present for both $\Delta m^2 = 0$ and $\Delta m^2 \ne 0$.
Interestingly,
it is known that focus points may be realized with any boundary 
condition of the form $(m^2_{H_u}, m^2_{U_3}, m^2_{Q_3})\propto(1,1+x,1-x)$ with $x$ an
arbitrary constant, the same form as those in this model.  Thus, we should expect
to obtain the same set of focus points in the model regardless of the choice of K\a"ahler moduli.
It should be emphasized that the more general choice of K\a"ahler moduli goes beyond 
mSUGRA/CMSSM. 
It is shown that 
there exists superpartner spectra with a light Higgsino-like LSP with $230-350$~GeV and a Higgs
mass in the range $124-126$~GeV, and which satisfy most standard experimental constraints.  
Consequently, viable spectra
with low EWFT between $3-7\%$ may be obtained.  The spin-independent direct-detection
crosssections are in range of future experiments such as XENON-1T and super CDMS, while
the relic density is smaller than the WMAP and Planck bounds by roughly a factor of ten,
implying that the LSP is sub-dominant component of dark matter.  In addition, 
most of the spectra are consistent with constraints from indirect-detection experiments.   
\end{abstract}


\maketitle
\thispagestyle{empty}

\newpage
\setcounter{page}{1}
\section{Introduction}

String theory is currently one of the most promising frameworks which permits a unification between 
quantum field theory and general relativity.  As such, it should be able to describe the physics 
of our universe in detail.  There are in fact many string theory compactifictions which
come very close to reproducing the Standard Model (SM) and its supersymmetric extension,
the Minimal Supersymmetric Standard Model (MSSM). 
Examples include heterotic string compactications
on orbifolds~\cite{Buchmuller:2005jr, Lebedev:2006kn, Kim:2006hw} 
and on Calabi-Yau manifolds~\cite{Braun:2005ux, Bouchard:2005ag},
and free fermionic models~\cite{AEHN, Faraggi:1989ka,
Antoniadis:1990hb, LNY, Cleaver:2001ab}.
More, recently compactifications involving D-branes have attracted much attention. 
These types of models fall into two general classes, intersecting/magnetized
D-brane models on orientifold backgrounds~\cite{Berkooz:1996km,
Ibanez:2001nd, Blumenhagen:2001te, CSU, Cvetic:2002pj, Cvetic:2004ui, Cvetic:2004nk, 
Cvetic:2005bn, Chen:2005ab, Chen:2005mj, Blumenhagen:2005mu}, and 
Gepner 
models~\cite{Dijkstra:2004ym, Dijkstra:2004cc}.  
One string-derived MSSM satisfying all global consistency conditions has been constructed
from intersecting/magnetized D-branes within the context 
of Type II orientifold compactifications~\cite{Mayes:2013bda,Chen:2007zu,Chen:2007px} 
on a $T^6/(\Z_2 \times \Z_2)$ 
background.  
This model contains three generations 
of quarks and leptons as well as a single pair of Higgs 
fields\footnote{Note that the Higgs sector of the model considered in~\cite{Mayes:2013bda} is different
than in~\cite{Chen:2007zu,Chen:2007px}}.  
The model contains a minimal amount of exotic matter, which may be decoupled
from the low-energy sector.  In addition, the tree-level gauge couplings are automatically unified
at the string scale~\cite{Chen:2007zu,Chen:2007px}. Thus, this is a phenomenologically
interesting model worthy of detailed study.  

Supersymmetry (SUSY) and supergravity (SUGRA) are intimately related to string theory. In particular, 
string theory requires supersymmetry for consistency.  Furthermore, supergravity theories arise
as the low-energy limits of string theories.  Although supersymmetry seems to be 
a required ingredient of string theory, this does not necessarily imply that it is 
broken at an energy scale such that we should expect supersymmetric partners
of the SM particles to be observed at energies accessible at the Large Hadron Collider (LHC).  
However, there are many phenomenological reasons to expect this.  In particular,
SUSY is one of the best-motivated solutions to the hierarchy problem.  
Although the exact mechanism and scale at which SUSY is broken in nature should it exist
is not known, simple calculations suggest that the masses of (at least some) of the superpartners should have
$\mathcal{O}(1$~TeV$)$ masses if SUSY solves the hierarchy problem without introducing any (or not too much) fine-tuning.  
Moreover, it can be shown that there is an upper bound on the Higgs mass
in the MSSM, $m_h \lesssim 130$~GeV~\cite{Carena:2002es}.
In addition to providing a solution to the hierarchy problem, 
SUSY with R-parity imposed can provide a natural candidate 
for dark matter~\cite{Ellis:1982wr,Ellis:1983wd,Ellis:1983ew}. Finally, 
the unification of the three gauge 
couplings when extrapolated to high energies via the Renormalization Group Equations (RGE) 
is much more precise when SUSY
is incorporated into the Minimal Supersymmetric Standard Model (MSSM) in comparison
to the non-SUSY SM, consistent with the idea of
Grand Unification Theories (GUTs)~\cite{Dimopoulos:1981yj, Ibanez:1981yh}.  

The discovery of a $124-126$~GeV 
Higgs boson~\cite{:2012gk,:2012gu} at the (LHC) has
so far not been followed by the discovery of any new physics which might explain the
hierarchy problem.  Supersymmetry remains one of the best candidates for such
new physics. 
However, direct searches during the first run of the LHC
for squarks and gluinos are pushing the mass limits for these particles into the 
TeV range~\cite{Aad:2012fqa,Aad:2011ib,Aad:2012hm,:2012mfa,Chatrchyan:2011zy}.   
Furthermore, to obtain a $\sim125$~GeV Higgs mass in the MSSM requires large
radiative corrections involving the top/stop sector, requiring large stop
squark masses $\mathcal{O}($TeV$)$ and/or large values of tan$\beta$.

The most-studied framework for supersymmetry breaking is 
minimal supergravity (mSUGRA), or 
equivalently the Constrained MSSM (CMSSM)~\cite{Chamseddine:1982jx, Ohta:1982wn, Hall:1983iz}.  
However, to obtain a sufficiently large Higgs mass in mSUGRA/CMSSM 
requires heavy squarks and sleptons which generically spoils the naturalness in which
the hierarchy problem is solved by introducing some amount of electroweak fine-tuning (EWFT).  One possible 
exception to this is the hyperbolic branch (HB)/focus point (FP) region of the mSUGRA/CMSSM parameter
space characterized by large $m_0$ in comparison to $m_{1/2}$ 
where the amount of required EWFT is minimized locally relative to the full parameter 
space with similar values of $m_0$~\cite{Chan:1997bi,Feng:1999mn,Feng:1999zg,Baer:1995nq,Baer:1998sz,Chattopadhyay:2003xi}.  
Several different groups have recently reassessed the status of mSUGRA/CMSSM
in light of the $\sim 125$~GeV Higgs 
discovery~\cite{Kadastik:2011aa,Strege:2011pk,Aparicio:2012iw,Ellis:2012aa,Baer:2012uya,Matchev:2012vf,Akula:2012kk,Ghosh:2012dh,Fowlie:2012im,Buchmueller:2012hv,Strege:2012bt,Citron:2012fg,Ellis:2012nv,Boehm:2012rh}  
A consensus has built that 
the mSUGRA/CMSSM parameter space is being strongly constrained by this discovery and pushed
into regions which require some degree of fine-tuning~\cite{Baer:2012mv}.  This has motivated
the study of extensions of mSUGRA/CMSSM such as the non-universal Higgs mass
models (NUHM) where the universality condition on the scalar soft masses 
of mSUGRA/CMSSM is relaxed such that the Higgs soft masses may be treated as independent 
parameters~\cite{Ellis:2002iu,Ellis:2002wv,Baer:2004fu,Baer:2005bu,Ellis:2012nv}.  
This allows superpartner spectra with a $\mu$-parameter close to the electroweak 
scale, thus featuring low fine-tuning, to be easily obtained. 
These models generically feature light Higgsinos with masses in the few-hundred GeV
range.  More generally, models which feature radiatively-driven natural supersymmetry 
(RNS) may be considered~\cite{Baer:2013xua}.

In the phenomenologically interesting intersecting/magnetized D-brane model discussed above, 
it is possible to study the possible
sets of supersupersymmetric soft terms which may be obtained, specifically for the F-type
supersymmetry breaking where SUSY-breaking is dominated by the complex structure and/or K\a"ahler
moduli as well as the universal dilaton.  In general, the sets of possible soft terms in the model are non-universal.  
Investigations into the sets of non-universal soft terms have been performed 
in~\cite{Chen:2007zu,Chen:2007px,Maxin:2009ez,Li:2014xqa}.
However, there also exists subsets of the soft terms which feature universality.  
As shown in~\cite{Mayes:2013bda}, these subsets 
are similar to the the effective mSUGRA/CMSSM parameter space in 
that they feature a universal scalar mass $m_0$, 
a universal gaugino mass $m_{1/2}$ and a
universal trilinear term $A_0$, which for the present case is fixed such that
$A_0 = -m_{1/2}$.  These boundary conditions coincide with the slice of the effective mSUGRA/CMSSM
parameter space studied in~\cite{Mayes:2013qmc}, which includes a band along the HB/FP branch
where spectra with a Higgs mass in the range $124-126$~GeV may be obtained in addition 
to sastifying experimental bounds on the dark matter relic density while also satisying 
most standard experimental constraints.  In addition, 
for generic values of the K\a"ahler moduli in the model, the 
soft terms for sfermions are split about the Higgs mass-squared terms as
$m_{Q_L,L_L}^2=m_H^2 - \Delta m^2$ and $m_{Q_R,L_R}^2=m_H^2 + \Delta m^2$ where $\Delta m^2$ is a function
of the K\a"ahler moduli.  Only for a specific choice of the K\a"ahler moduli do the soft terms
reduce to those with a universal scalar mass for both Higgs and sfermions.
Interestingly,
it was observed in~\cite{Feng:1999zg} that focus points may be realized with any boundary 
condition of the form $(m^2_{H_u}, m^2_{U_3}, m^2_{Q_3})\propto(1,1+x,1-x)$ with $x$ an
arbitrary constant.  These are essentially the same as boundary conditions 
in the model that we are considering, the only difference being that the soft terms
in the model have this form for all three families of sfermions. Thus, we should expect
to obtain the same set of focus points in the model regardless of the choice of K\a"ahler moduli.
It should be emphasized that the more general choice of K\a"ahler moduli goes beyond 
mSUGRA/CMSSM.

In the following, detailed scans of 
the D-brane model supersymetry parameter space with $A_0=-m_{1/2}$ have been performed.  
Regions of this parameter space featuring light Higgsinos, a $\sim 125$~GeV Higgs mass, and low EWFT
are identified and discussed.  In particular, it is found that there 
exists superpartner spectra with a light Higgsino-like LSP with $225-350$~GeV masses and a Higgs
mass in the range $124-126$~GeV, and which satisfy most standard experimental constraints.  
Consequently, viable spectra
with low EWFT between $3-7\%$ may be obtained.  The dark matter direct-detection
cross-sections for these spectra are in range of future experiments such as 
XENON-1T~\cite{Aprile:2012zx} and super CDMS~\cite{Akerib:2006rr}.  In addition, 
most of the spectra are consistent with constraints from indirect-detection experiments.
Finally, the parameter space where the scalar mass-squared
terms for squarks and sleptons are split about the Higgs mass-squared term,
$m_{Q_L,L_L}^2=m_H^2 - \Delta m^2$ and $m_{Q_R,L_R}^2=m_H^2 + \Delta m^2$, 
is also studied in detail. As mentioned, this parameter space is obtained from the string model
for generic values of the  K\a"ahler moduli and goes beyond 
mSUGRA/CMSSM.   As expected, the HB/FP regions are also present for 
this region of the parameter space, independent of the choice of K\a"ahler moduli.

\section{High Scale Boundary Conditions}

Let us consider the realistic MSSM constructed from interesecting/magnetized D-branes 
studied in~\cite{Mayes:2013bda}. This simple and elegant model has many desirable phenomenological features
such as three families of quarks and leptons, minimal exotic matter, gauge coupling unification,
and rank-3 Yukawa mass matrices~\cite{Chen:2007px,Chen:2007zu} such that it is possible to obtain the 
correct fermion masses and mixings.  In addition, one variation of the model allows 
either baryon or lepton number to exists as a local gauge symmetry, thus 
forbidding proton decay~\cite{Maxin:2011ne}.  The model also satisfies 
all conditions for global consistency such as tadpole and anomaly cancellation. 

Type II orientifold string compactifications with intersecting/magnetized 
D-branes  
have provided useful geometric tools with which the MSSM may
be engineered~\cite{Blumenhagen:2005mu,Blumenhagen:2006ci}.  
To briefly give an over view of the construction of such models, D6-branes in Type IIA fill
(3+1)-dimensional Minkowski spacetime and wrap 3-cycles in the
compactified manifold, such that a stack of $N$ branes generates a
gauge group U($N$) [or U($N/2$) in the case of $T^6/(\Z_2 \times
\Z_2)$] in its world volume.  
In general, the 3-cycles wrapped by the stacks of D6-branes intersect
multiple times in the internal space, resulting
in a chiral fermion in the bifundamental representation localized at
the intersection between different stacks $a$ and $b$.  The multiplicity of such
fermions is then given by the number of times the 3-cycles intersect.
Each stack of D6-branes $a$ may 
intersect the orientifold images of other stacks $b'$, also resulting in fermions in
bifundamental representations.  Each stack may also intersect its own
image $a'$, resulting in chiral fermions in the symmetric and
antisymmetric representations.  In addition, the
consistency of the model requires certain constraints to be satisfied,
namely, Ramond-Ramond (R-R) tadpole cancellation and the preservation 
of $\mathcal{N}=1$ supersymmetry.

The model discussed in~\cite{Mayes:2013bda} describes a 
three-generation Pati-Salam model with additional hidden sectors.  The
full gauge symmetry of the model is given by $[{\rm U}(4)_C \times {\rm U}(2)_L \times {\rm
U}(2)_R]_{\rm observable} \times [ {\rm USp}(2)^4]_{\rm hidden}$.
As discussed in detail
in~\cite{Chen:2007px,Chen:2007zu}, with this configuration of D6 branes all R-R
tadpoles are canceled, K-theory constraints are satisfied, and
$\mathcal{N}=1$ supersymmetry is preserved.  
Using the effective scalar potential it is
possible to study the stability ~\cite{Blumenhagen:2001te}, the
tree-level gauge couplings \cite{CLS1, Shiu:1998pa,Cremades:2002te}, 
gauge threshold corrections \cite{Lust:2003ky},
and gauge coupling unification \cite{Antoniadis:Blumen}.  The
effective Yukawa couplings \cite{Cremades:2003qj, Cvetic:2003ch},
matter field K\"ahler metric and soft-SUSY breaking terms have
also been investigated \cite{Kors:2003wf}.  A more detailed
discussion of the K\"ahler metric and string scattering of gauge,
matter, and moduli fields has been performed in
\cite{Lust:2004cx}.  

The $\mathcal{N}=1$ supergravity action depends upon three
functions, the holomorphic gauge kinetic function, $f$, K\a"ahler
potential $K$, and the superpotential $W$.  Each of these will in
turn depend upon the moduli fields which describe the background
upon which the model is constructed.
Supersymmetry is broken when some of the F-terms of the hidden sector fields $M$
acquire VEVs. This then results in soft terms being generated in
the observable sector. For simplicity, it is assumed in this
analysis that the $D$-term does not contribute (see
\cite{Kawamura:1996ex}) to the SUSY breaking.  Then, the goldstino
is eaten by the gravitino via the superHiggs effect and 
thus the gravitino obtains a mass, 
$m_{3/2}$. 

The
normalized gaugino mass parameters, scalar mass-squared
parameters, and trilinear parameters respectively may be given in
terms of the K\a"ahler potential, the gauge kinetic function, and
the superpotential as
\begin{eqnarray}
M_P &=& \frac{1}{2\mbox{Re}f_P}(F^M\partial_M f_P), \\ \nonumber
m^2_{PQ} &=& (m^2_{3/2} + V_0) - \sum_{M,N}\bar{F}^{\bar{M}}F^N\partial_{\bar{M}}\partial_{N}log(\tilde{K}_{PQ}), \\ \nonumber
A_{PQR} &=& F^M\left[\hat{K}_M + \partial_M log(Y_{PQR}) - \partial_M log(\tilde{K}_{PQ}\tilde{K}_{QR}\tilde{K}_{RP})\right],
\label{softterms}
\end{eqnarray}
where $\tilde{K}_{QR}$ is the K\a"ahler metric appropriate for D-branes
which are parallel on at least one torus, i.e. involving
non-chiral matter.
We assume that the supersymmetric breaking in the model
is dominated by the complex structure moduli $u^i$, $i=1,2,3$ which describe the shape of the compactified dimensions
as well as the universal dilaton $s$. 
We allow the dilaton $s$ and the $u$-moduli to obtain vacuum expectation values (VEVs).
To do this, we parameterize the $F$-terms as
\begin{equation}
F^{u^i,s} = \sqrt{3}m_{3/2}[(s + \bar{s})\Theta_s  + (u^i + \bar{u}^i)\Theta_i^u + (t^i + \bar{t}^i)\Theta_i^t]
\end{equation}
The parameters $\Theta_i$ parameterize the goldstino direction
in $U^i$ space,  where $\sum (|\Theta_i^u|^2  + |\Theta_s|^2 =1$. The
goldstino angle $\Theta_s$ determines the degree to which SUSY
breaking is being dominated by the dilaton $s$ and/or complex
structure ($u^i$) and K\a"ahler ($t^i$) moduli. For the present, we do not 
allow the K\a"ahler moduli $t^i$
to obtain VEVs so that $\Theta^t_i = 0$.     

In general, the soft terms for intersecting D-brane models are non-universal.  However,
universal soft terms may arise in the model under study for the case $\Theta^u_1 = \Theta^u_2 \equiv \Theta_{12}$ 
and $\Theta^u_3 = \Theta_S \equiv \Theta_{3s}$.  We refer the reader to ref.~\cite{Mayes:2013bda} for a detailed derivation
of the resulting soft terms.  
For convenience, let us define
$\Theta_{12} = \frac{1}{\sqrt{2}}~\mbox{cos}\theta$ and $\Theta_{3s}=\frac{1}{\sqrt{2}}~\mbox{sin}\theta$.  
Then, the soft terms take the simple form
\begin{eqnarray}
m^2_H = \frac{m^2_{3/2}}{4},
\end{eqnarray}
\begin{eqnarray}
m_{1/2} = \sqrt{\frac{3}{2}}m_H\left[1+\mbox{sin}(2\theta) \right]^{1/2},
\end{eqnarray}
\begin{eqnarray}
m^2_{Q_L, L_L} = m^2_H - \Delta m^2,
\end{eqnarray}
\begin{eqnarray}
m^2_{Q_R, L_R} = m^2_H + \Delta m^2,
\end{eqnarray}
\begin{eqnarray}
A_0 =-m_{1/2}, 
\label{softterms2}
\end{eqnarray}
with
\begin{eqnarray}
\Delta m^2 = \frac{6m^2_H}{\pi}\left[\mbox{cos}^2\theta - \mbox{sin}^2\theta\right]\psi(t,\bar t),
\end{eqnarray}
where $\psi(t,\bar t)$ is a function which depends on the K\a"ahler moduli, as shown in~\cite{Mayes:2013bda} 
and $m_{3/2}$ is the gravitino mass.
\begin{figure}
  \centering
	\includegraphics[width=1.0\textwidth]{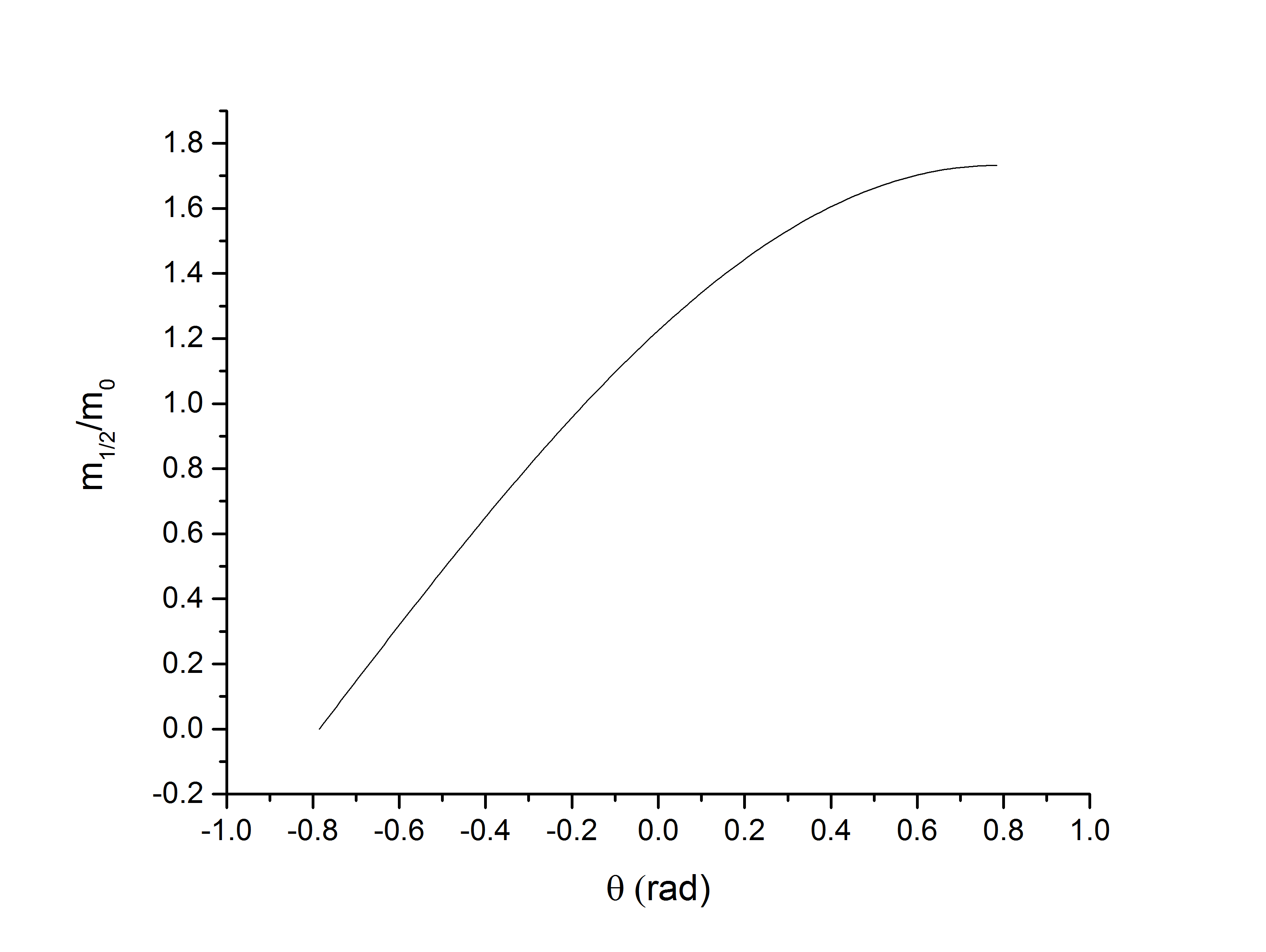}
			\caption{Gaugino mass to scalar mass ratio as a function of $\theta$ for $\Delta m^2=0$, i.e. mSUGRA/CMSSM with $A_0=-m_{1/2}$.}
	\label{fig:GauginoScalarMassvsTheta.png}
\end{figure}

\begin{figure}
  \centering
	\includegraphics[width=1.0\textwidth]{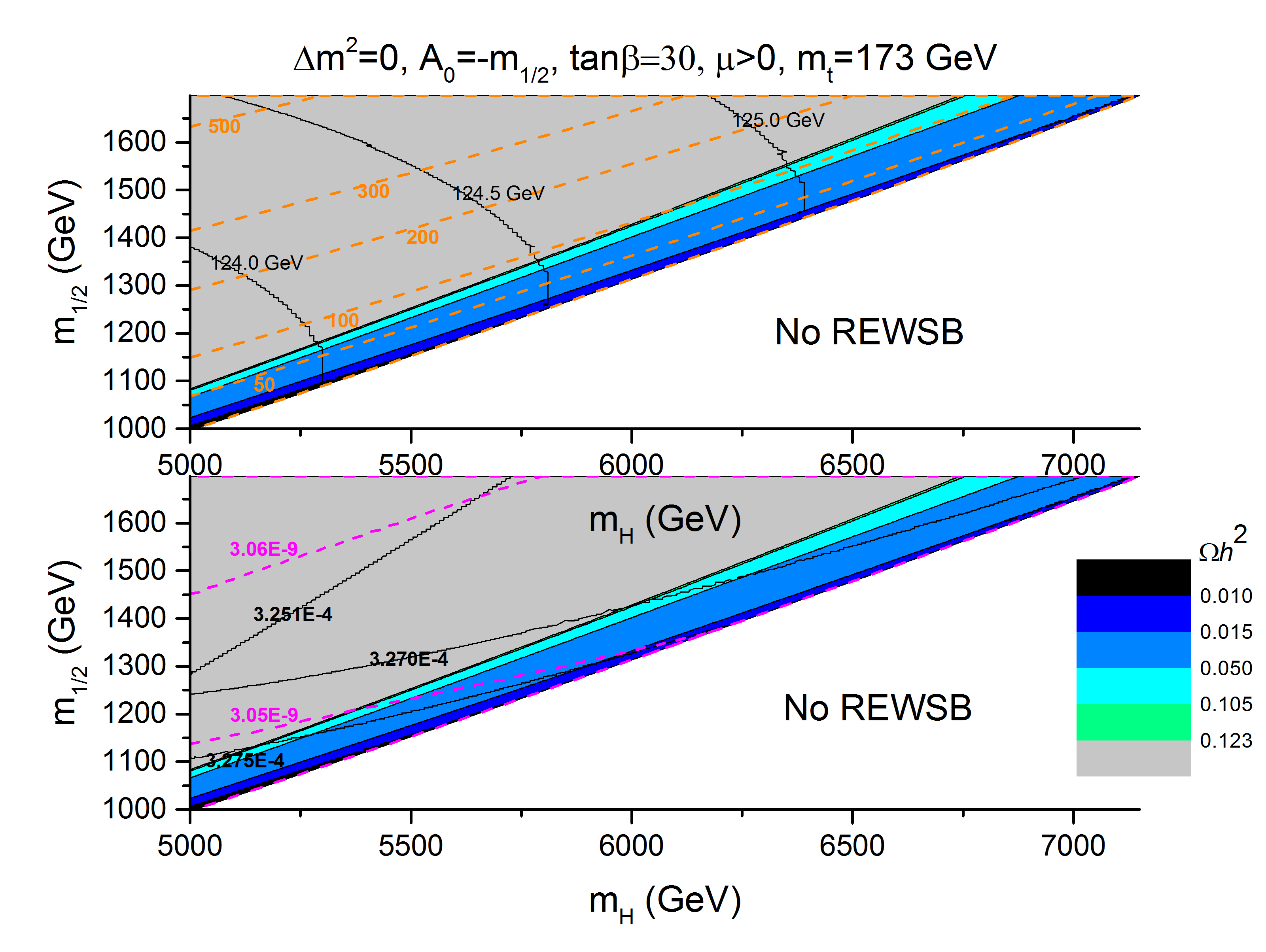}
			\caption{The $m_{1/2} \ vs.\ m_H=m_0$ plane with $\Delta m^2=0$, $A_0 = -m_{1/2}$, $\mu>0$, tan$\beta=30$, and $m_t=173$~GeV.  The regions shaded in dark blue indicate $\Omega_{\chi^0} h^2 \lesssim 0.123$, while the region shaded in green satisfies the WMAP 9-year 2$\sigma$ bounds, $0.105 \lesssim \Omega_{\chi^0} h^2 \lesssim 0.123$.  Top panel: Values of the higgs mass $m_h$ are denoted by solid black contours while dashed orange contours indicate values for minimal electroweak fine-tuning, 
			$\Delta_{EW}$.  Bottom panel: Values for the Flavor Changing Neutral Current (FCNC) process, $b \rightarrow s\gamma$ are indicated by solid black contours, while 
dashed magenta contours indicate values for the process $B_{s}^{0} \rightarrow \mu^+ \mu^-$.} 
	\label{fig:mSUGRA_CountourPlanetb30}
\end{figure}
  
\begin{figure}	
  \centering
	\includegraphics[width=1.0\textwidth]{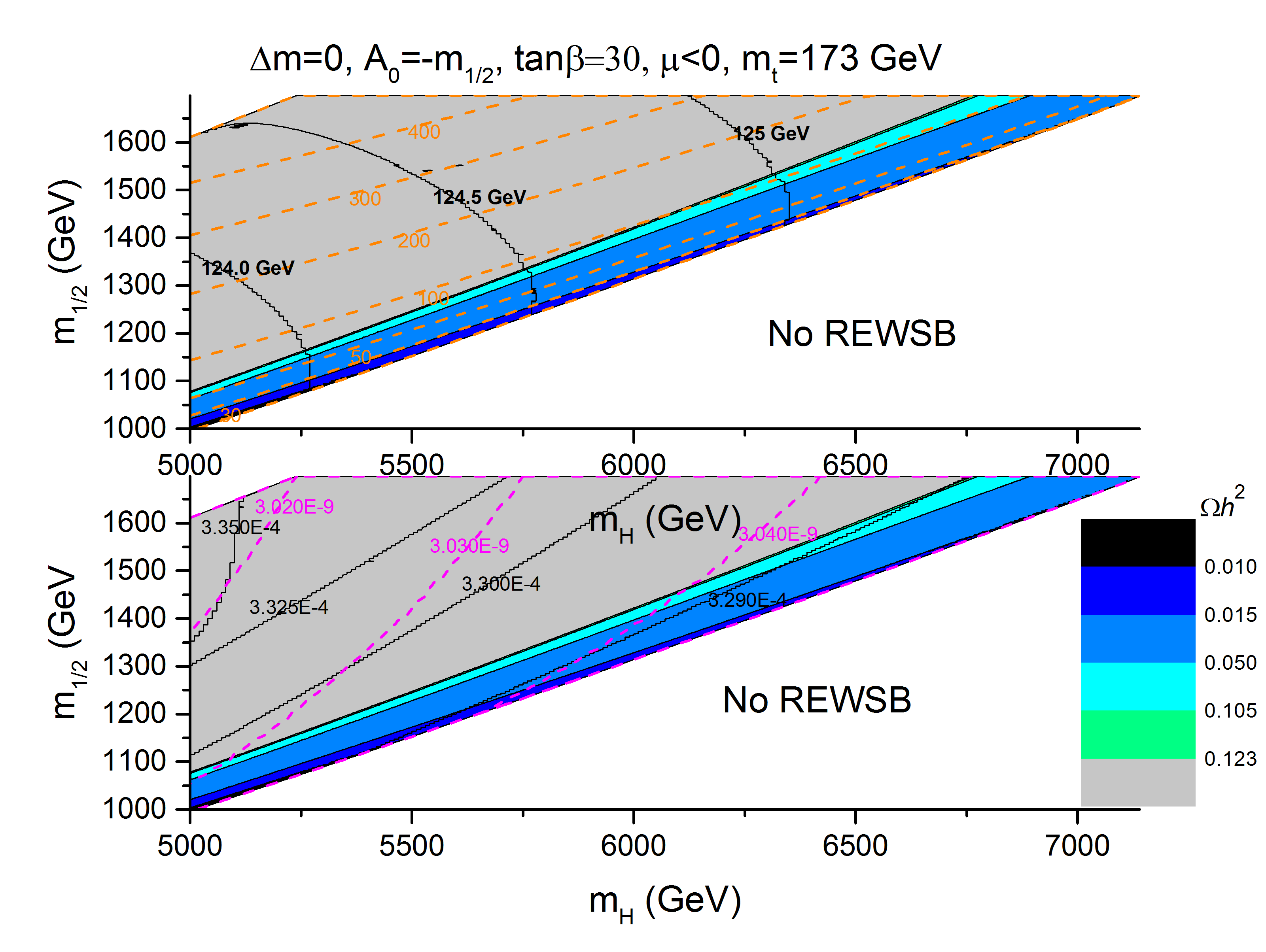}
			\caption{The $m_{1/2} \ vs.\ m_H=m_0$ plane with $\Delta m^2=0$, $A_0 = -m_{1/2}$, $\mu<0$, tan$\beta=30$, and $m_t=173$~GeV.  The regions shaded in dark blue indicate $\Omega_{\chi^0} h^2 \lesssim 0.123$, while the region shaded in green satisfies the WMAP 9-year 2$\sigma$ bounds, $0.105 \lesssim \Omega_{\chi^0} h^2 \lesssim 0.123$.  Top panel: Values of the higgs mass $m_h$ are denoted by solid black contours while dashed orange contours indicate values for minimal electroweak fine-tuning, 
			$\Delta_{EW}$.  Bottom panel: Values for the Flavor Changing Neutral Current (FCNC) process, $b \rightarrow s\gamma$ are indicated by solid black contours, while 
dashed magenta contours indicate values for the process $B_{s}^{0} \rightarrow \mu^+ \mu^-$.}
	\label{fig:mSUGRA_CountourPlanetb30muneg}
\end{figure}

\begin{figure}
  \centering
	\includegraphics[width=1.0\textwidth]{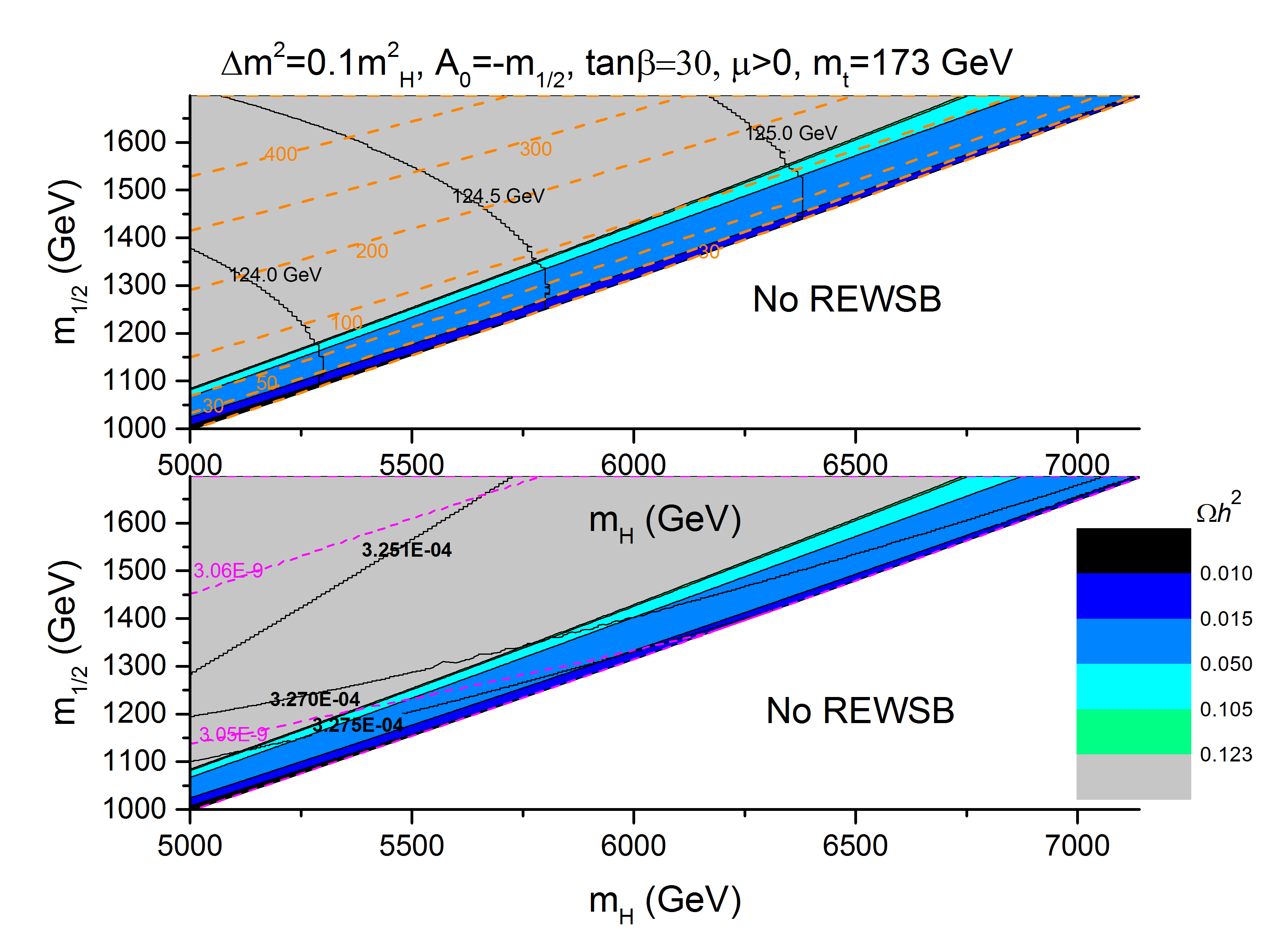}
			\caption{The $m_{1/2} \ vs.\ m_H=m_0$ plane with $\Delta m^2=0.1m^2_H$, $A_0 = -m_{1/2}$, $\mu>0$, tan$\beta=30$, and $m_t=173$~GeV.  The regions shaded in dark blue indicate $\Omega_{\chi^0} h^2 \lesssim 0.123$, while the region shaded in green satisfies the WMAP 9-year 2$\sigma$ bounds, $0.105 \lesssim \Omega_{\chi^0} h^2 \lesssim 0.123$.  Top panel: Values of the higgs mass $m_h$ are denoted by solid black contours while dashed orange contours indicate values for minimal electroweak fine-tuning, 
			$\Delta_{EW}$.  Bottom panel: Values for the Flavor Changing Neutral Current (FCNC) process, $b \rightarrow s\gamma$ are indicated by solid black contours, while 
dashed magenta contours indicate values for the process $B_{s}^{0} \rightarrow \mu^+ \mu^-$.}
	\label{fig:mSUGRA_CountourPlanetb30mupossplit1}
\end{figure}
  
\begin{figure}	
  \centering
	\includegraphics[width=1.0\textwidth]{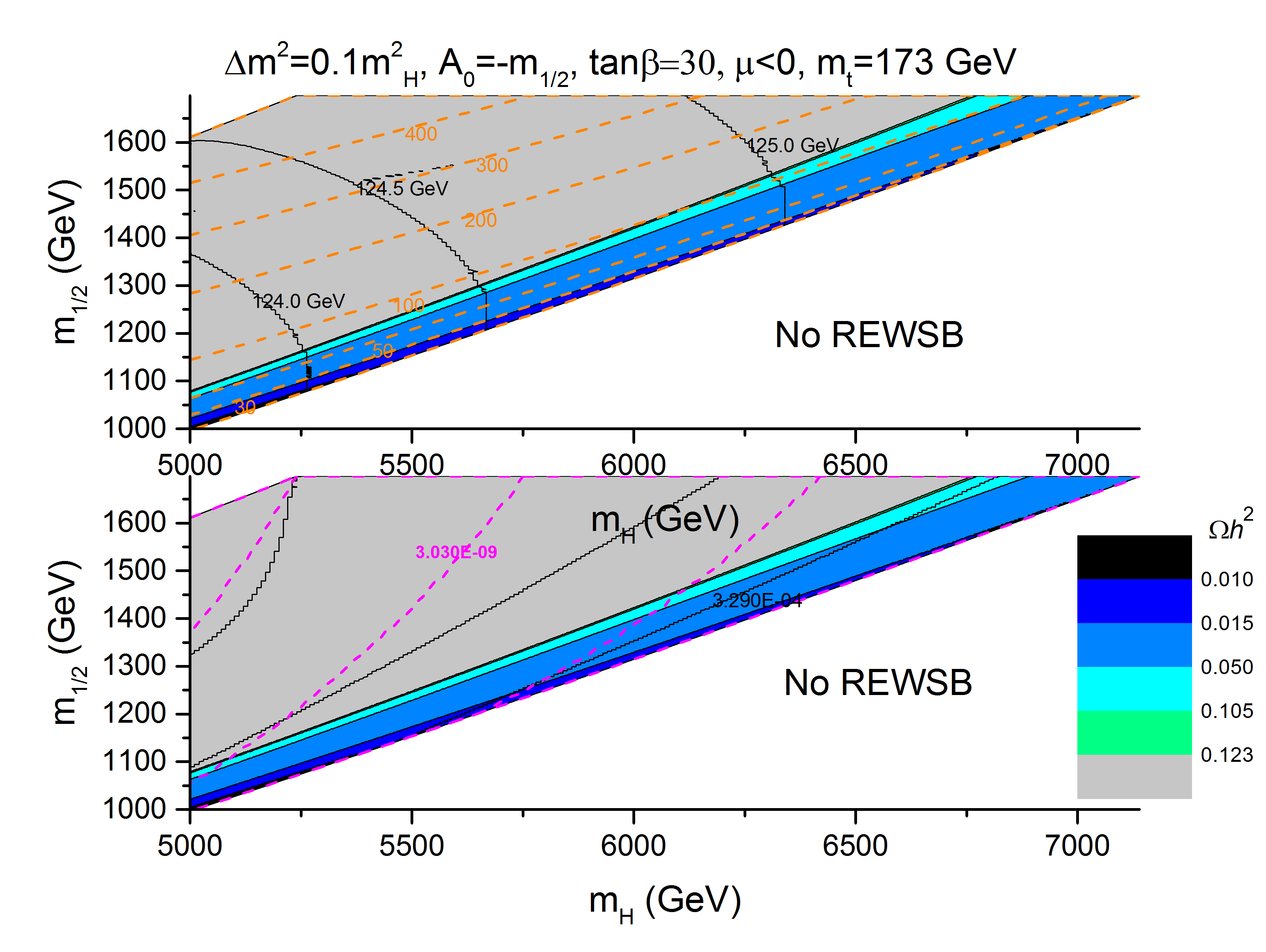}
			\caption{The $m_{1/2} \ vs.\ m_H=m_0$ plane with $\Delta m^2=0.1m^2_H$, $A_0 = -m_{1/2}$, $\mu<0$, tan$\beta=30$, and $m_t=173$~GeV.  The regions shaded in dark blue indicate $\Omega_{\chi^0} h^2 \lesssim 0.123$, while the region shaded in green satisfies the WMAP 9-year 2$\sigma$ bounds, $0.105 \lesssim \Omega_{\chi^0} h^2 \lesssim 0.123$.  Top panel: Values of the higgs mass $m_h$ are denoted by solid black contours while dashed orange contours indicate values for minimal electroweak fine-tuning, 
			$\Delta_{EW}$.  Bottom panel: Values for the Flavor Changing Neutral Current (FCNC) process, $b \rightarrow s\gamma$ are indicated by solid black contours, while 
dashed magenta contours indicate values for the process $B_{s}^{0} \rightarrow \mu^+ \mu^-$.}
	\label{fig:mSUGRA_CountourPlanetb30mupossplit2}
\end{figure}	

\begin{figure}	
  \centering
	\includegraphics[width=1.0\textwidth]{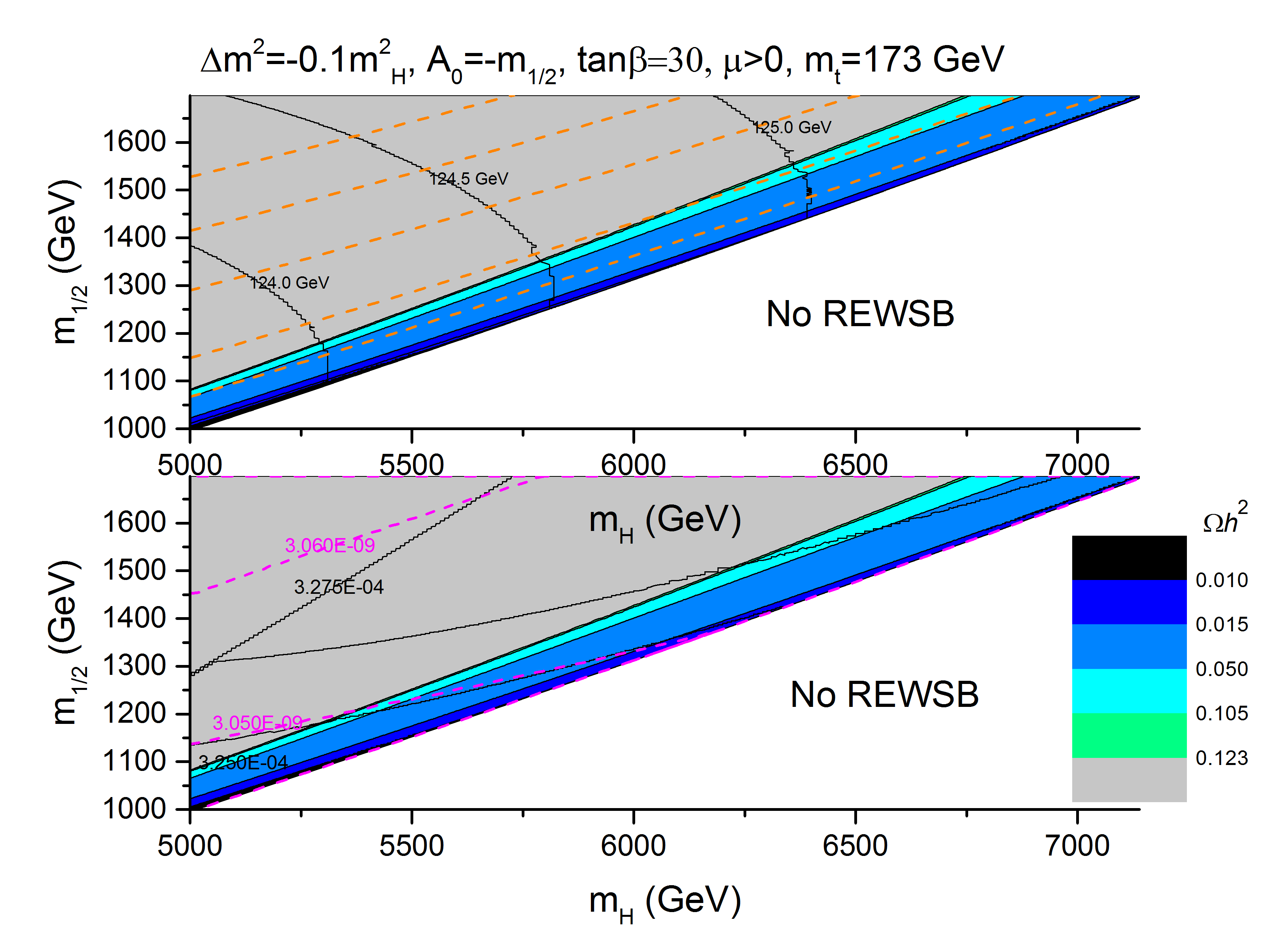}
			\caption{The $m_{1/2} \ vs.\ m_H=m_0$ plane with $\Delta m^2=-0.1m^2_H$, $A_0 = -m_{1/2}$, $\mu>0$, tan$\beta=30$, and $m_t=173$~GeV.  The regions shaded in dark blue indicate $\Omega_{\chi^0} h^2 \lesssim 0.123$, while the region shaded in green satisfies the WMAP 9-year 2$\sigma$ bounds, $0.105 \lesssim \Omega_{\chi^0} h^2 \lesssim 0.123$.  Top panel: Values of the higgs mass $m_h$ are denoted by solid black contours while dashed orange contours indicate values for minimal electroweak fine-tuning, 
			$\Delta_{EW}$.  Bottom panel: Values for the Flavor Changing Neutral Current (FCNC) process, $b \rightarrow s\gamma$ are indicated by solid black contours, while 
dashed magenta contours indicate values for the process $B_{s}^{0} \rightarrow \mu^+ \mu^-$.}
	\label{fig:mSUGRA_CountourPlanetb30mupossplit3}
\end{figure}	

\begin{figure}	
  \centering
	\includegraphics[width=1.0\textwidth]{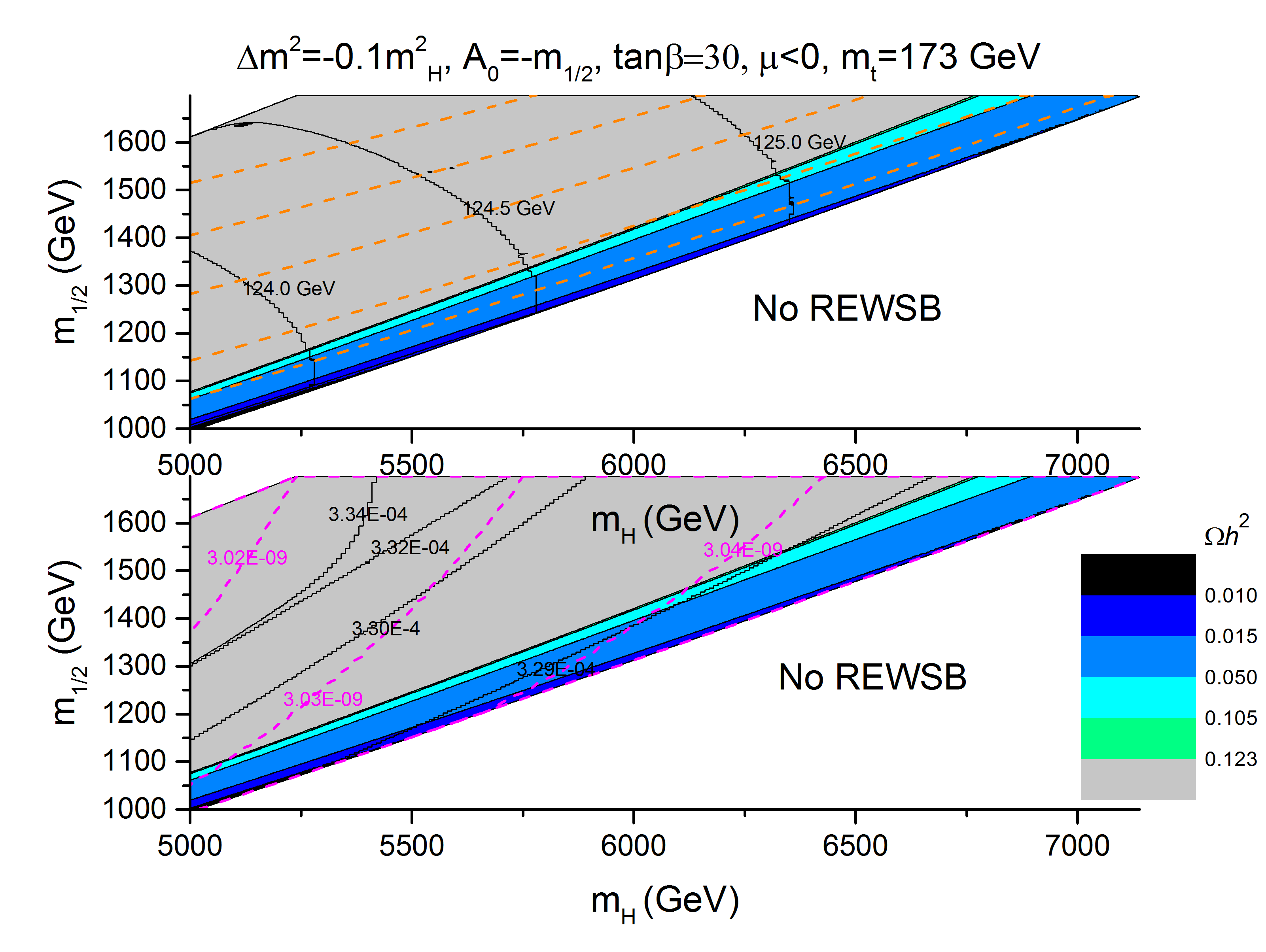}
			\caption{The $m_{1/2} \ vs.\ m_H=m_0$ plane with $\Delta m^2=-0.1m^2_H$, $A_0 = -m_{1/2}$, $\mu<0$, tan$\beta=30$, and $m_t=173$~GeV.  The regions shaded in dark blue indicate $\Omega_{\chi^0} h^2 \lesssim 0.123$, while the region shaded in green satisfies the WMAP 9-year 2$\sigma$ bounds, $0.105 \lesssim \Omega_{\chi^0} h^2 \lesssim 0.123$.  Top panel: Values of the higgs mass $m_h$ are denoted by solid black contours while dashed orange contours indicate values for minimal electroweak fine-tuning, 
			$\Delta_{EW}$.  Bottom panel: Values for the Flavor Changing Neutral Current (FCNC) process, $b \rightarrow s\gamma$ are indicated by solid black contours, while 
dashed magenta contours indicate values for the process $B_{s}^{0} \rightarrow \mu^+ \mu^-$.}
	\label{fig:mSUGRA_CountourPlanetb30mupossplit4}
\end{figure}

There are two cases of special note where $\Delta m^2=0$, resulting in a universal scalar mass.  
First, taking $\theta = \pi/4$ the soft terms
become those of the {\it special dilaton} solution, 
\begin{eqnarray}
m_{1/2}=\frac{\sqrt{3}}{2}m_{3/2}, \ \ \ \ \ \ \ \ \ m_0 = \frac{m_{3/2}}{2}, \ \ \ \ \ \ \ \ A_0 = -m_{1/2},
\end{eqnarray}
which is well-known from heterotic string compactifications. In fact, these boundary conditions are
generic for all Pati-Salam models constructed from intersecting D-branes as shown in~\cite{Mayes:2013bda}. 
A recent study
of this parameter space has been performed~\cite{Maxin:2008kp,Maxin:2009pr}.  
The second case where $\Delta m^2=0$ is for a specific value of the K\a"ahler moduli as shown 
in~\cite{Mayes:2013bda}.  The parameter
space for this scenario essentially is similar to the case of mSUGRA/CMSSM in that it features a universal scalar mass $m_0$, a universal gaugino mass $m_{1/2}$ and a universal trilinear term which is fixed as
$A_0 = -m_{1/2}$ for the present case.    
For more general values of the K\a"ahler moduli, the scalar masses-squared for sfermions are split about
the Higgs scalar mass-squared.  In principle, it should be possible to dynamically fix the K\a"ahler moduli, complex structure moduli
and the dilaton in the model by including non-perturbative effects such as supergravity fluxes or gaugino condensation in a hidden sector.  
However, for the present
we shall treat the K\a"ahler moduli as free parameters, and there  is no gaurantee that they will take the special case noted above where 
$\Delta m^2 = 0$.  Our objective in the following sections is thus to study the more general parameter space with $\Delta m^2 \ne 0$,
which occurs for generic values of the K\a"ahler moduli, in addition to the case with $\Delta m^2 = 0$.  

As noted in the Introduction, it was observed in~\cite{Feng:1999zg} that focus points may be realized with any boundary 
condition of the form $(m^2_{H_u}, m^2_{U_3}, m^2_{Q_3})\propto(1,1+x,1-x)$ with $x$ an
arbitrary constant.  The above soft terms have exactly this form with $\Delta m^2 = x$.  
Thus, we should find that a focus point in the parameter space with $\Delta m^2 = 0$ is also a focus point 
when $\Delta m^2 \ne 0$.  Since $\Delta m^2$ is a function which depends on the 
K\a"ahler moduli, it should be the case that the HB/FP region of the parameter space is essentially independent
of the K\a"ahler moduli.  Indeed this is the case, as we shall see.

\section{Parameter Space}

As noted in the previous section, the  high-scale boundary conditions for this model are universal
for a specific value of the K\a"ahler moduli.  For this case the soft terms are similar to those
of the Constrained Minimal Supersymmetric Standard Model (CMSSM),  
one of the most studied model of supersymmetry breaking.  These models are characterized by the following 
parameters: a universal scalar mass $m_0$, a universal gaugino mass $m_{1/2}$, the 
Higgsino mixing $\mu$-parameter, the Higgs bilinear $B$-parameter, a universal 
trilinear coupling $A_0$, and tan~$\beta$.  One then determines the $B$ and $|\mu|$ 
parameters by the minimization of the Higgs potential triggering 
REWSB~\cite{Ellis:1983bp,AlvarezGaume:1983gj}, with the sign 
of $\mu$ remaining undetermined.  One is then left with the following free parameters
at the GUT scale: $m_0$, $m_{1/2}$, $A_0$, tan$\beta$ and sign($\mu$). 
For the string-derived model of the previous section, we may also trade $B$ and $\mu$ for 
tan$\beta$ and sign($\mu$) so that we have effectively the same set of GUT scale
parameters as for the CMSSM, with the trilinear terms related to the gaugino mass as
$A_0 = -m_{1/2}$.  However, it should be kept in mind that the truly fundamental parameters 
are $m_{1/2}$, $\theta$, $B$ and $\mu$ from which the effective parameters $m_0$, $m_{1/2}$, 
and $A_0$ are related by Eqns.~\ref{softterms2}.  In addition, the sfermion mass splitting
$\Delta m$ depends upon the  K\a"ahler moduli of the string-derived model.  This splitting
is zero for a specific value of the K\a"ahler moduli, and so the GUT-scale parameters are
universal for this case.  In the following, we will study the resulting supersymmetry parameter
space for both $\Delta m = 0$ and $\Delta m \ne 0$.

The high-scale boundary conditions derived from the model
are input into
{\tt SuSpect 2.41}~\cite{SUSP} to evolve the soft terms down to the electroweak scale via 
the Renormalization Group Equations (RGEs) and then input into  
{\tt MicrOMEGAs 3.6.9.2}~\cite{Belanger:2010pz,Belanger:2008sj,Belanger:2006is}to 
calculate the corresponding relic neutralino density, direct-detection cross-sections, and
indirect-detection annihilation cross-sections.  
In particular, we shall focus on the FP/HB
region of the parameter space.   
We vary $m_0$ (or equivalently $m_H$ for the cases with split sfermion masses) in increments of $2$~GeV between 
$5000-7500$~GeV and $m_{1/2}$ in increments of $1$~GeV for each scan.  
We fix tan$\beta=30$ for convenience, although our results are generic to other values of tan$\beta$.  
The Higgsino mass parameter
$\mu$ is determined by the requirement of 
radiative electroweak symmetry breaking (REWSB).  In addition to imposing experimental constraints,
the spectra are filtered from the final data set if the iterative procedure employed  by {\tt SuSpect} 
does not converge to a reliable solution.  
We take the top quark mass to be
$m_t = 173.3 \pm 1.1$~GeV~\cite{ATLAS:2014wva}.   We do not fix the sign of $\mu$ as the contribution to 
$g_{\mu}-2$ for the muon is small for superpartner spectra with heavy scalars.  
In analyzing the resulting data, we consider the following experimental constraints:

\begin{enumerate}

\item The first results from the Planck experiment\cite{Ade:2013lta}, with a dark matter density in the range $\Omega_c h^2 = 0.1199 \pm 0.0027$
and
the WMAP 9-year $2\sigma$ preferred range~\cite{Hinshaw:2012fq} for the cold dark matter density,  0.105 $\leq \Omega_{\chi^o} h^{2} \leq$ 0.123.   We consider two cases, one where a neutralino LSP is the dominant component of the dark matter and another where it makes up a subdominant component such that
0 $\leq \Omega_{\chi^o} h^{2} \leq$ 0.123.

\item The experimental limits on the Flavor Changing Neutral Current (FCNC) process, $b \rightarrow s\gamma$. The results from the Heavy Flavor Averaging Group (HFAG)~\cite{HFAG}, in addition to the BABAR, Belle, and CLEO results, are: $Br(b \rightarrow s\gamma) = (355 \pm 24^{+9}_{-10} \pm 3) \times 10^{-6}$. There is also a more recent estimate~\cite{MMEA} of $Br(b \rightarrow s\gamma) = (3.15 \pm 0.23) \times 10^{-4}$. For our analysis, we use the limits $2.86 \times 10^{-4} \leq Br(b \rightarrow s\gamma) \leq 4.18 \times 10^{-4}$, where experimental and
theoretical errors are added in quadrature.

\item The process $B_{s}^{0} \rightarrow \mu^+ \mu^-$ which has recently been observed 
      to be in the range $2\times 10^{-9} < BF(B_{s}^{0} \rightarrow \mu^+ \mu^-) < 4.7\times 10^{-9}$ by LHCb~\cite{:2012ct}.

\item The lightest CP-even Higgs mass in the range $124$~GeV$\lesssim m_h \lesssim 126$~GeV as observed by the ATLAS and CMS experiments at the LHC~\cite{:2012gk,:2012gu}.

\end{enumerate}

We will not require that the problem of the anomalous magnetic moment of the muon~\cite{MUON}, $4.7\times 10^{-10} \leq a_{\mu} \approx 52.7\times 10^{-10}$, is solved by 
contributions from supersymmetric particles as the spectra that will be studied may only make small contributions.  Furtheremore, there are large hadronic contributions
to this anomaly that require delicate subtractions with large uncertainties~\cite{Davier:2010nc}.

\begin{figure}	
  \centering
	\includegraphics[width=1.0\textwidth]{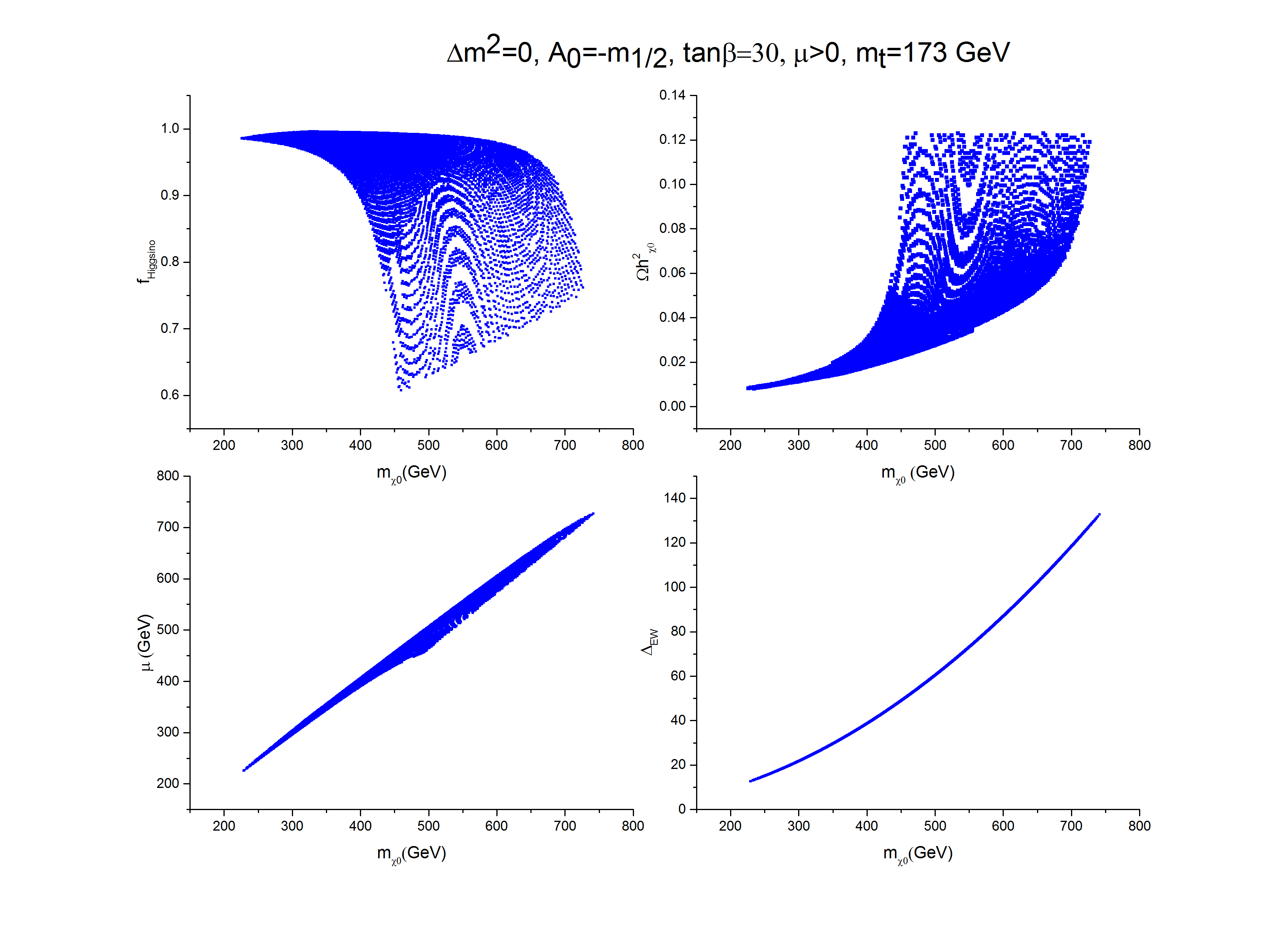}
			\caption{Top left panel: Higgsino fraction vs. LSP mass. Top right panel: Relic density vs. LSP mass.
			         Bottom left panel: Higgsino bilinear $\mu$ vs. LSP mass $m_{\tilde{\chi}^0}$.
							 Bottom right panel:  Minimal electroweak fine-tuning $\Delta_{EW}$ vs. LSP mass.  
							 For each plot, $m_{\tilde{\chi}^0}$ for $\Delta m^2=0$, tan$\beta=30$, $\mu>0$, and $m_t=173$~GeV.}  
	\label{fig:DMdensityvsNeutralinoMasstb30mupos}
\end{figure}	

\begin{figure}	
  \centering
	\includegraphics[width=1.0\textwidth]{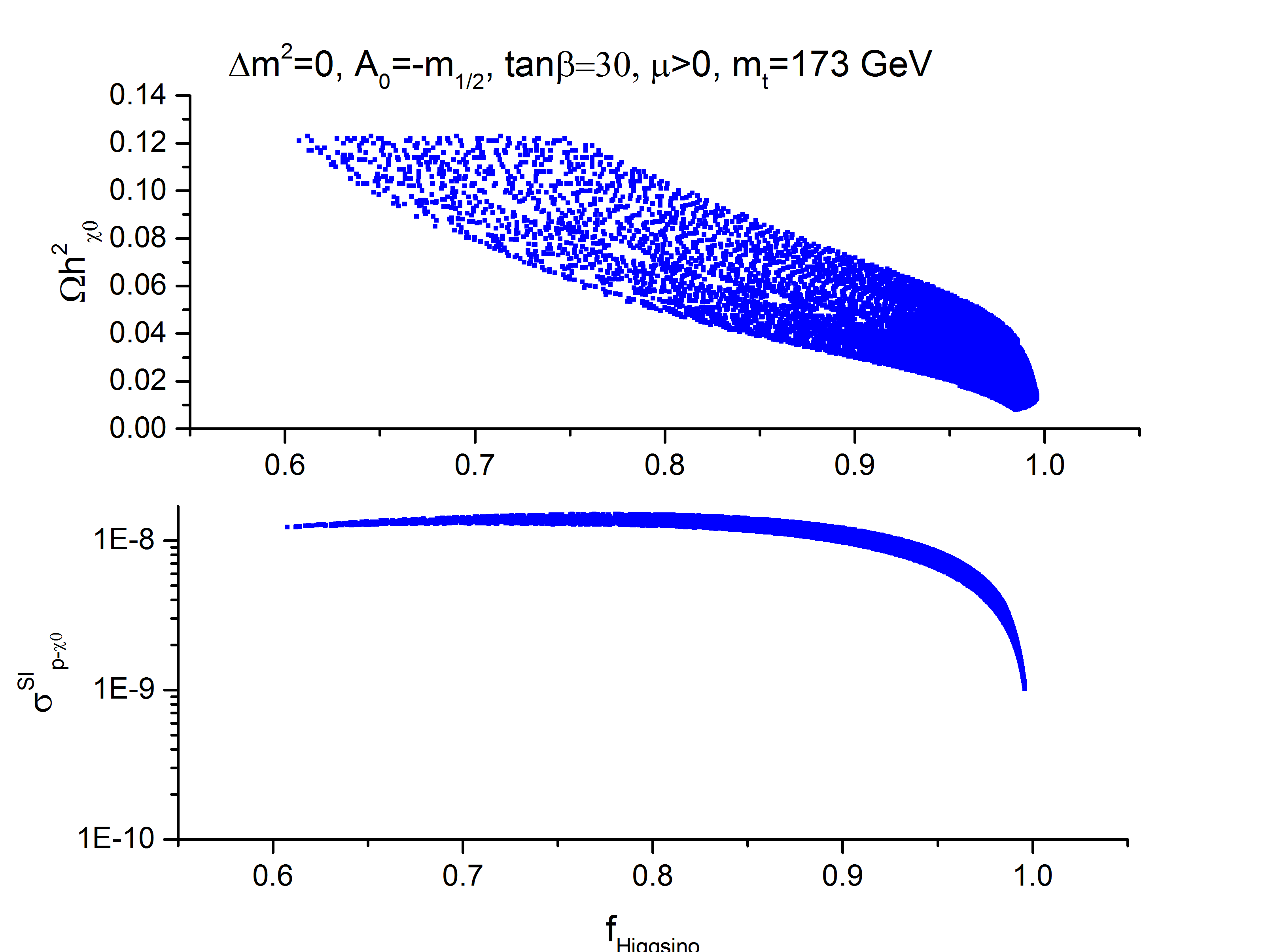}
			\caption{Top panel: LSP relic density $\Omega_{\tilde{\chi}^0}$ vs. higgsino fraction $f_{higgsino}$ for $\Delta m^2=0$, tan$\beta=30$, $\mu>0$, and
			 $m_t=173$~GeV. Bottom panel: Neutralino-proton spin-independent direct-detection cross-sections $\sigma^{SI}_{\tilde{\chi}^0}$ vs. higgsino fraction $f_{Higgsino}$
		for $\Delta m^2=0$, tan$\beta=30$ and $\mu>0$.}
	\label{fig:DMdensityvsHiggsinoFractiontb30mupos}
\end{figure}

\begin{figure}	
  \centering
	\includegraphics[width=1.0\textwidth]{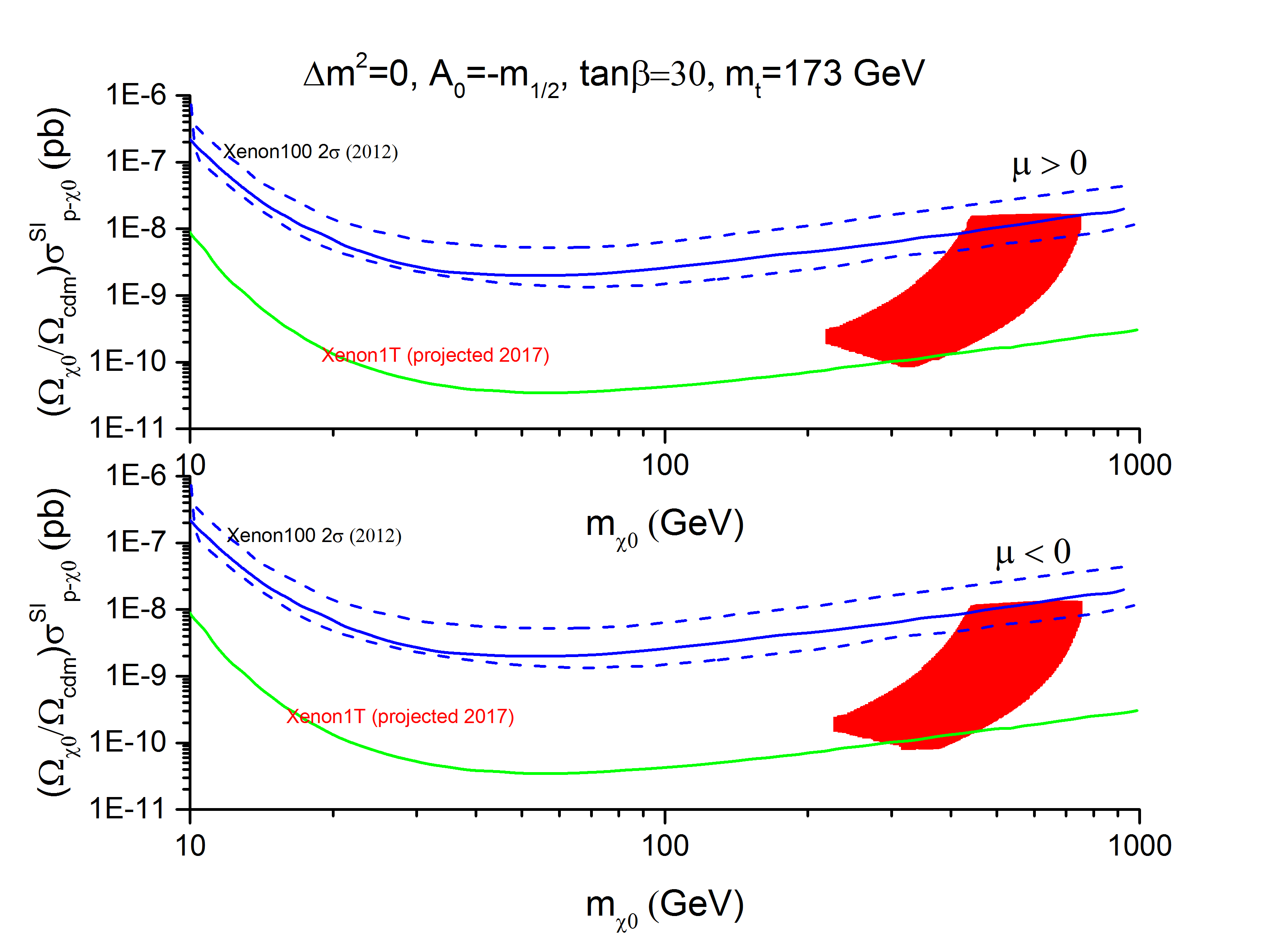}
			\caption{Re-scaled neutralino spin-independent direct-detection cross-sections $(\Omega_{\tilde{\chi}^0}/\Omega_{DM})\times\sigma^{SI}_{\tilde{\chi}^0}$vs. 
			LSP mass $m_{\tilde{\chi}^0}$ for $\Delta m^2=0$, tan$\beta=30$, and $m_t=173$~GeV. Here, we take $\Omega_{DM}h^2=0.11$. The top panel is for $mu>0$ while
			the bottom panel shows $m<0$.}
	\label{fig:DirectDetectCrossSectionstb30mupos}
\end{figure}	

\begin{figure}	
  \centering
	\includegraphics[width=1.0\textwidth]{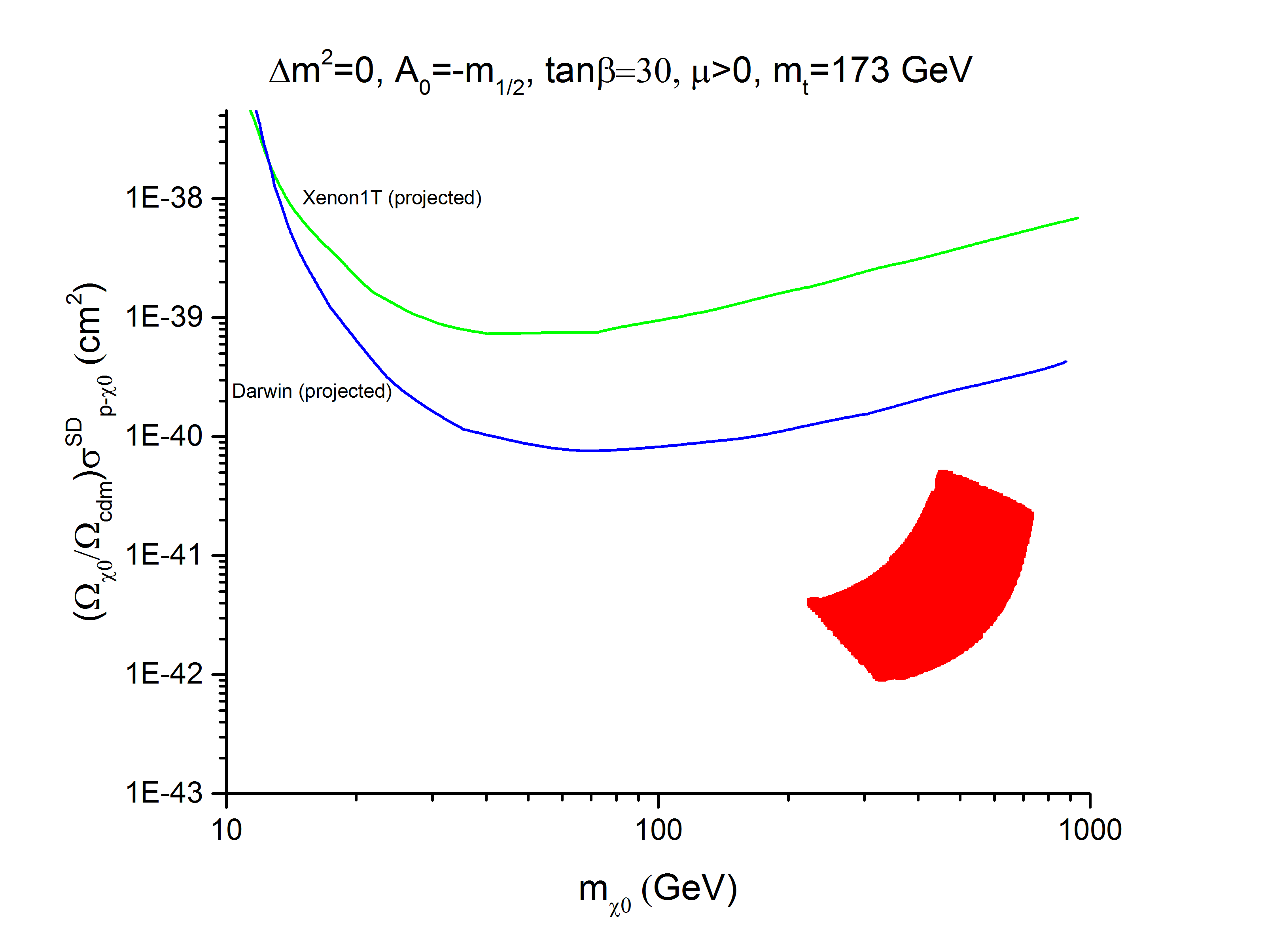}
			\caption{Re-scaled neutralino spin-dependent direct-detection cross-sections $(\Omega_{\tilde{\chi}^0}/\Omega_{DM})\times\sigma^{SD}_{\tilde{\chi}^0}$vs. 
			LSP mass $m_{\tilde{\chi}^0}$ for $\Delta m^2=0$, tan$\beta=30$, $\mu>0$, and $m_t=173$~GeV. Here, we take $\Omega_{DM}h^2=0.11$. Also shown are 
			the projected reach for the Xenon1T~\cite{Garny:2012it} and Darwin~\cite{Baudis:2012bc} dark matter direct-detection experiments.}
	\label{fig:SDDirectDetectCrossSectionstb30mupos}
\end{figure}

\begin{figure}	
  \centering
	\includegraphics[width=1.0\textwidth]{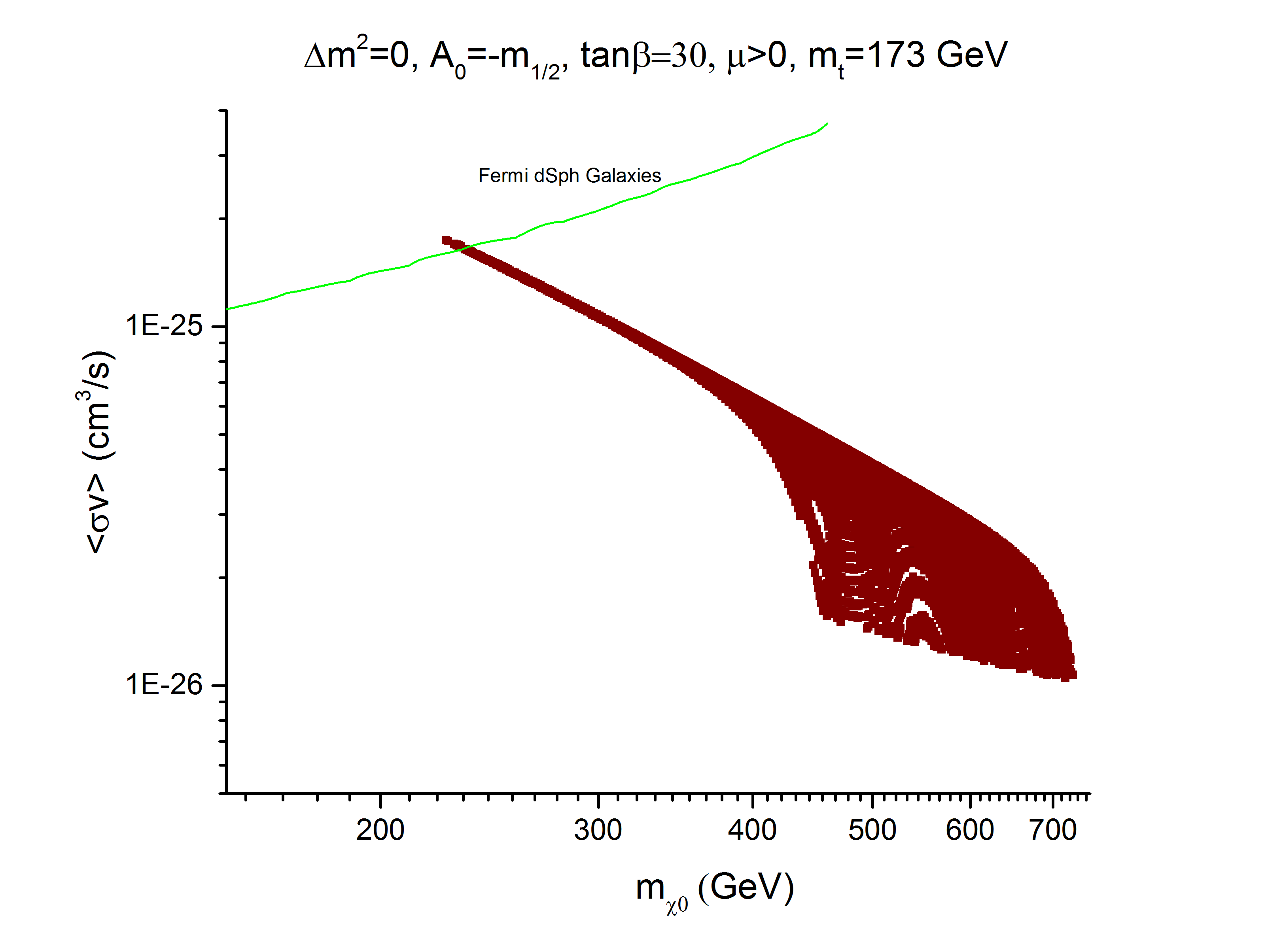}
			\caption{Annihilation cross-sections to continuum photons $<\sigma v>$ vs. LSP neutralino mass $m_{\tilde{\chi}_0}$ for model points satisfying 
			 $\Omega_{\chi_0}h^2 \lesssim 0.123$ for $\Delta m^2=0$, tan$\beta=30$, $\mu>0$, and $m_t=173$~GeV.}
	\label{fig:AnnihilationCrossSections}
\end{figure}

\begin{table}[t]
\footnotesize
\renewcommand{\arraystretch}{1.0}
\caption{Benchmark points for for $\Delta m^2=0$, tan$\beta=30$, $\mu>0$, $m_t=173$~GeV, and $\Delta_{EW} \lesssim 30$ .}
\label{Spectrum}
\begin{center}
\begin{tabular}{|c|c|c|c|c|c|c|c|c|}\hline
$m_{0}$~(GeV) & $m_{1/2}$~(GeV) & $m_h$~(GeV) &$m_{\tilde{\chi}^0}$~(GeV)  & $m_{\tilde{g}}$~(TeV) & $m_{\tilde{t}_1}$~(TeV) & $\Omega_{\chi_0}h^2$ & $\sigma_{\tilde{\chi_0}}^{SI}$~(pb) & $\Delta_{EW}$\\
\hline\hline
$5200$ & $1059.0$ & $124.0$ & $236.4$ & $2.58$ & $3.48$ & $8.6\times 10^{-3}$ & $3.2\times 10^{-9}$  &  $13.7$\\
$5500$ & $1153.5$ & $124.3$ & $253.3$ & $2.79$ & $3.71$ & $9.3\times 10^{-3}$ & $2.6\times 10^{-9}$  &  $15.6$\\
$5750$ & $1233.0$ & $124.5$ & $263.5$ & $2.96$ & $3.90$ & $9.7\times 10^{-3}$ & $2.2\times 10^{-9}$  &  $16.7$\\
$6000$ & $1313.5$ & $124.7$ & $276.4$ & $3.13$ & $4.09$ & $1.0\times 10^{-2}$ & $1.9\times 10^{-9}$  &  $18.3$\\
$6250$ & $1397.3$ & $124.9$ & $294.5$ & $3.31$ & $4.28$ & $1.1\times 10^{-2}$ & $1.7\times 10^{-9}$  &  $20.6$\\
$6500$ & $1479.0$ & $125.1$ & $301.4$ & $3.49$ & $4.47$ & $1.2\times 10^{-2}$ & $1.4\times 10^{-9}$  &  $21.9$\\
$6750$ & $1562.0$ & $125.3$ & $311.1$ & $3.67$ & $4.66$ & $1.2\times 10^{-2}$ & $1.2\times 10^{-9}$  &  $22.8$\\
$7000$ & $1646.5$ & $125.5$ & $320.5$ & $3.85$ & $4.85$ & $1.3\times 10^{-2}$ & $1.1\times 10^{-9}$  &  $24.1$\\ 
$7250$ & $1732.0$ & $125.7$ & $332.5$ & $4.03$ & $5.05$ & $1.3\times 10^{-2}$ & $9.7\times 10^{-10}$ &  $25.9$\\
$7500$ & $1818.0$ & $125.9$ & $342.5$ & $4.21$ & $5.24$ & $1.4\times 10^{-2}$ & $8.6\times 10^{-10}$ &  $27.3$\\
$7750$ & $1905.0$ & $126.0$ & $350.1$ & $4.39$ & $5.44$ & $1.5\times 10^{-2}$ & $7.6\times 10^{-10}$ &  $28.5$\\
$8000$ & $1992.5$ & $126.2$ & $358.2$ & $4.57$ & $5.63$ & $1.5\times 10^{-2}$ & $6.7\times 10^{-10}$ &  $29.8$\\
\hline
\end{tabular}
\end{center}
\end{table}

\begin{table}[t]
\footnotesize
\renewcommand{\arraystretch}{1.0}
\caption{Benchmark points for for $\Delta m^2=0$, tan$\beta=30$, $\mu>0$, $m_t=173.3$~GeV, and $0.105 \leq \Omega_{\chi_0} h^2 \geq 0.123$.}
\label{Spectrum2}
\begin{center}
\begin{tabular}{|c|c|c|c|c|c|c|c|c|}\hline
$m_{0}$~(GeV) & $m_{1/2}$~(GeV) & $m_h$~(GeV) &$m_{\tilde{\chi}^0}$~(GeV)  & $m_{\tilde{g}}$~(TeV) & $m_{\tilde{t}_1}$~(TeV) & $\Omega_{\chi_0}h^2$ & $\sigma_{\tilde{\chi_0}}^{SI}$~(pb) & $\Delta_{EW}$\\
\hline\hline
$5000$ & $1082.0$ & $123.7$ & $454.7$ & $2.26$ & $3.39$ & $1.1\times 10^{-1}$ & $1.3\times 10^{-8}$ &  $58.1$\\
$5190$ & $1146.0$ & $123.9$ & $483.3$ & $2.76$ & $3.54$ & $1.1\times 10^{-1}$ & $1.3\times 10^{-8}$ &  $64.6$\\
$5510$ & $1257.0$ & $124.3$ & $533.7$ & $3.00$ & $3.79$ & $1.2\times 10^{-1}$ & $1.3\times 10^{-8}$ &  $77.0$\\
$5740$ & $1336.0$ & $124.5$ & $567.8$ & $3.17$ & $3.97$ & $1.2\times 10^{-1}$ & $1.3\times 10^{-8}$ &  $85.5$\\
$6000$ & $1427.0$ & $124.7$ & $607.5$ & $3.36$ & $4.17$ & $1.1\times 10^{-1}$ & $1.3\times 10^{-8}$ &  $96.1$\\
$6250$ & $1518.0$ & $125.0$ & $649.0$ & $3.56$ & $4.37$ & $1.2\times 10^{-1}$ & $1.3\times 10^{-8}$ &  $108.6$\\
$6500$ & $1606.0$ & $125.2$ & $685.2$ & $3.74$ & $4.56$ & $1.1\times 10^{-1}$ & $1.3\times 10^{-8}$ &  $118.9$\\
$6750$ & $1697.0$ & $125.4$ & $723.6$ & $3.93$ & $4.76$ & $1.1\times 10^{-1}$ & $1.3\times 10^{-8}$ &  $130.9$\\
\hline
\end{tabular}
\end{center}
\end{table}

One of the most compelling motivations for introducing low-scale SUSY is to solve the hierarchy problem.  
An important question is whether or not this is
accomplished naturally without reintroducing any fine-tuning.  
In fact, the mass of the observed Higgs boson is 
slightly too large to accomodate in models such as mSUGRA/CMSSM without some level of fine-tuning. 
This is refererred to as the \textit{little hierarchy problem}.  Ordinarily, spectra with large scalar 
masses would generically be considered highly fine-tuned.  This is not necessarily true for those spectra which fall 
in the HB/FP region of the parameter space, which may feature reduced electroweak fine-tuning relative to the local parameter
space with similar values of $m_0$. 
The amount of fine-tuning with respect to the electroweak scale (EWFT) is typically 
quantified by the minimal electroweak fine-tuning parameter 
\begin{equation}
\Delta_{EW} \equiv \mbox{max}(C_i)/(M_Z^2/2),
\end{equation}
where $C_{\mu}\equiv \left|-\mu^2\right|$, $C_{H_u}\equiv\left|-m^2_{H_u}tan^2\beta/(tan^2\beta-1)\right|$, and 
$C_{H_d}\equiv \left|-m^2_{H_d}/(tan^2\beta-1)\right|$.  The percent-level of EWFT is then given by $\Delta_{EW}^{-1}$.
It should be noted that for most of the parameter space explored in this analysis $C_{\mu}$ is dominant, 
and so generally we have
$\Delta_{EW}=\left|-\mu^2\right|/(M_Z^2/2)$. In the following, we shall use this parameter to analyze the minimal
amount of fine-tuning required regions of the studied parameter space. 

In~\cite{Baer:2012mv}, it is argued that $\Delta_{EW}$ only provides a measure of
the minimum amount of fine-tuning in regards to the electroweak scale and provides no information about the high
scale physics involved in a particular model of SUSY breaking.  In order to provide a measure of how fine-tuned a particular
model is given knowledge of how SUSY is broken at a high energy scale, a parameter called $\Delta_{HS}$ was introduced which
is analogous to $\Delta_{EW}$~\cite{Baer:2012mv}.  For most of the parameter space this parameter is given by
\begin{equation}
\Delta_{HS} = \frac{m_0^2+\mu^2}{(M_Z^2/2)}= \Delta_{EW}+\frac{m_0^2}{(M_Z^2/2)}.
\end{equation}
As we can see, for regions of the mSUGRA/CMSSM parameter space with large scalar masses, $\Delta_{HS}$ is
very large even for those cases where $\Delta_{EW}$ is small such as in the HB regions.  This simply reflects
the fact that, although a particular SUSY spectrum may be completely natural and solve the hierarchy problem
without any fine-tuning, obtaining this spectrum within the mSUGRA/CMSSM framework of SUSY breaking 
requires large cancellations which only happens for specific sets of soft-terms rather than the general parameter 
space.  For the specific string model at hand, we will assume that high-scale parameters $m_{3/2}$ and $\theta$, and thus 
$m_{1/2}$ and $m_0$, are
chosen by whatever dynamics are responsible for moduli stabilization, thus at least providing an explanation for the choice 
of high-scale parameters.  We leave this topic for future work.

Contour plots of the $m_{1/2} \ vs. \ m_H=m_0$ plane with $\Delta m^2=0$ for tan$\beta=30$ are shown for $\mu >0$ and $\mu <0$ 
respectively in Figs.~\ref{fig:mSUGRA_CountourPlanetb30}-\ref{fig:mSUGRA_CountourPlanetb30muneg}.  The mass
of the lightest CP-even Higgs is between $124-126$~GeV for the regions shown. The viable areas of the parameter space 
lie along the HB/FP regions of the parameter space where $m_0=m_H\sim(4-5)\times m_{1/2}$.  As can be seen from 
Fig.~\ref{fig:GauginoScalarMassvsTheta.eps}, these regions can be obtained from the intersecting/magnetized D-brane 
model for $\theta\approx-(0.6-0.7)$~rad.
The dark blue regions in the figures have a relic density which 
satisfies $\Omega_{\tilde{\chi_0}} h^2 <=0.0123$, the light blue regions satisfy $0.015 <= \Omega_{\tilde{\chi_0}} h^2 <= 0.05$,
while the regions in green satisfy $0.105 <= \Omega_{\tilde{\chi_0}} h^2 <= 0.123$ corresponding to the 9-year 
$2\sigma$ bounds on the
dark matter
density observed by WMAP.  The gray regions have a relic density larger than the WMAP and Planck bounds.  
The upper panel of the plot shows values of the Higgs mass indicated by solid black contours and are in the range
$124-126$~GeV, while values of the electroweak
fine-tuning parameter $\Delta_{EW}$ are indicated by dashed orange contours.  For the parameter space with 
$\Omega_{\tilde{\chi_0}} h^2 <=0.0123$, $\Delta_{EW} \lesssim 130$.  
The bottom panel of the plot indicates values for the Flavor Changing Neutral Current (FCNC) process, 
$b \rightarrow s\gamma$ by solid black contours and are in the range $(3.27-3.30)\times 10^{-4}$, while 
dashed magenta contours indicate values for the process $B_{s}^{0} \rightarrow \mu^+ \mu^-$ and are in the range 
$(3.03-3.04)\times 10^{-9}$ consistent with experimental constraints. 
As can be observed, the parameter spaces for both signs of $\mu$ are very similar.  In addition, it should be noted
that the parameter space for different values of tan$\beta$ are also very similar, although here we have only shown
the parameter space for tan$\beta=30$.  In addition, varying the top quark mass within its experimental uncertainty 
shifts the parameter space slightly in the $m_{1/2} \ vs. \ m_H=m_0$ plane, however essentially the same results are obtained.
Please note that although these plots seem to indicate that these spectra lie along a continuous band, they are actually
interspersed with spectra where {\tt SuSpect} is not able to converge to a solution.  

Plots of the parameter space with $\Delta m^2 = 0.1m_H^2$ are shown in 
Figs.~\ref{fig:mSUGRA_CountourPlanetb30mupossplit1} 
and~\ref{fig:mSUGRA_CountourPlanetb30mupossplit2} for tan$\beta=30$
with $\mu>0$ and $\mu<0$ respectively, 
while 
plots of the parameter space with $\Delta m^2 = -0.1m_H^2$ are shown in 
Figs.~\ref{fig:mSUGRA_CountourPlanetb30mupossplit3} 
and~\ref{fig:mSUGRA_CountourPlanetb30mupossplit4} for tan$\beta=30$
with $\mu>0$ and $\mu<0$ respectively.
As we can see from these plots, the HB/FP regions are still present even when the sfermion
mass-squared terms are split about the Higgs scalar mass-squared term.  
Similar results are obtained for varying levels of splitting. 
This is, of course, expected
since it was previously observed that any boundary conditions of the form 
$(m^2_{H_u}, m^2_{U_3}, m^2_{Q_3})\propto(1,1+x,1-x)$ with $x$ an
arbitrary constant may give rise to focus points.  In the D-brane model at hand,
the amount of splitting depends upon the K\a"ahler moduli as well as on the Goldstino
angle $\theta$.  These results demonstrate that the HB/FP region is present 
for arbitrary values of the K\a"ahler moduli.  

A plot of the relic density vs. LSP mass is 
shown in the top-right panel of Fig.~\ref{fig:DMdensityvsNeutralinoMasstb30mupos} for $\Delta m^2 =0$, tan$\beta=30$, and $\mu>0$.  
Here, we can see that the LSP mass
is in the range $\approx 450-750$~GeV for superpartner spectra where the relic density falls within the WMAP $2\sigma$ bounds,
$0.105 <= \Omega_{\tilde{\chi_0}} h^2 <= 0.123$.
Also from this plot, we can see that when the LSP mass is in the range $240-350$~GeV, the relic density
is minimum, $\Omega_{\tilde{\chi_0}} h^2 \approx 0.010-0.015$.  A plot of the higgsino fraction which composes
the LSP vs. the LSP mass is shown in the upper-left panel of Fig.~\ref{fig:DMdensityvsNeutralinoMasstb30mupos}.  
From this plot, it can be seen that for LSP's with masses
in the range $240-350$~GeV are almost pure higgsino.  

For a scenario with a mostly higgsino LSP which composes roughly $10\%$ of the observed dark matter 
density, it has been suggested that the bulk of the dark matter is composed of axions~\cite{Bae:2013bva}. 
In the present context, superheavy hidden sector states may also provide some component of the dark matter.
Such states tend to generically carry fractional electric charges, however they may be confined
into neutral bound states by the hidden sector gauge interaction~\cite{Ellis:1990iu}, 
similar to how quarks become bound into hadrons.  It is known that superheavy particles
$X$ with masses in the range $10^{11} \lesssim M_X \lesssim 10^{14}$ might well have been produced 
naturally through the interaction of the vacuum with the gravitational field during the reheating 
period of the Universe following inflation in numbers sufficient to produce 
superheavy dark matter~\cite{Kolb}. For the model under study, it was shown that the hidden sector states become confined at high 
energy scales $10^{11}-10^{12}$~GeV~\cite{Chen:2007zu}, just in the preferred range.  
Thus, these bound states are a natural candidate for dark matter, in addition to the neutralino 
LSP. 

In the upper panel of Fig.~\ref{fig:DMdensityvsHiggsinoFractiontb30mupos}, a strong
correlation between the relic density and the higgsino fraction $f_{higgsino}$ of the linear combination of states
composing the LSP may be observed.  It can be seen from this plot
that the miminal relic density occurs for an almost pure higgsino LSP. It may be inferred from this that 
the higher the higgsino fraction composing the LSP, the lighter the LSP mass. This result is, of course, not unexpected.
For spectra satisfying  
$0.105 <= \Omega_{\tilde{\chi_0}} h^2 <= 0.123$, the higgsino fraction is in the range $0.65-0.85$ with the rest being
primarily of bino composition.  In general, the higher the bino fraction, the larger the relic density and the larger 
the LSP mass.  In addition, there is a vanishingly small wino fraction.  

The lower-left panel of Fig.~\ref{fig:DMdensityvsNeutralinoMasstb30mupos} shows a plot of $\mu$ vs. LSP mass, from which is can be seen that
$m_{\tilde{\chi_0}}\approx \mu$.  
The lower-right panel of Fig.~\ref{fig:DMdensityvsNeutralinoMasstb30mupos} displays a plot of the minimal electroweak fine-tuning parameter
$\Delta_{EW}$ vs. LSP mass.  From this plot it can be seen that $\Delta_{EW} \sim (50-120)$ 
for superpartner spectra where $0.105 <= \Omega_{\tilde{\chi_0}} h^2 <= 0.123$.  Thus, such spectra are minimally fine-tuned at 
the level of $\Delta_{EW}^{-1}=1-2\%$.  For those spectra with a light higgsino-like LSP with a mass in the range $230-350$~GeV,
the minimal fine-tuning is much less, in the range of $\Delta_{EW}^{-1}=3-7\%$.  This result can generally be understood 
given that the $\mu$-parameter is essentially equivalent to the higgsino mass as well as being a measure of the minimal 
electreak fine-tuning, viz-\a'a-viz $C_{\mu}\equiv \left|-\mu^2\right|$.  

For all of the spectra for which the Higgs mass satisfies $124$~GeV$\lesssim m_h \lesssim 126$~GeV
and for which the relic density satisfies the WMAP and Planck constraints, the gluino mass is in the range $2.6-4.4$~TeV, while
the lightest stop mass is in the range $3.5-5.6$~TeV.  The squarks and sleptons
all have masses greater than the gluino mass. Sets of benchmark points are shown in Table~\ref{Spectrum} and Table~\ref{Spectrum2} spanning a range of Higgs masses
from $124-126$~GeV for the cases $\Delta_{EW}\lesssim 30$ and $0.105 \leq \Omega_{\chi_0} h^2 \geq 0.123$ respectively.   
In particular, only the four lightest neutralinos and charginos have masses which are 
below $1$~TeV, with a small mass splitting of $5-10$~GeV between the LSP and NLSP.  These spectra are very similar
to those obtained in RNS models with a light higgsino LSP.  The possible signals that may be observed for these spectra have
been considered in~\cite{Baer:2013xua}.  Due to the heavy masses for the gluino and squarks in these models, it would be
very difficult to observe a clear and distinct signal at the LHC if the spectrum of superpartners falls into these regions of the
parameter space.  Direct production of light higgsinos may produce clear, soft trilepton signatures if the mass gap between
the LSP and NLSP is not too great.  
The prospects for observing superpartners at a linear collider or at a higher-energy hadron collider 
appear to be somewhat more promising~\cite{Baer:2011ec}.  
In the case of the linear collider, this is because 
lightest neutralinos and charginos may be produced even with a lower CM energy and their
decays may be studied with cleaner backgrounds, whereas it may be possible to directly produce 
stops and gluinos at a hadron collider with higher collision energy.

Although it may be difficult to observe a distinct signal at the LHC if nature has chosen one of these spectra, 
the situation is somewhat better in the case of direct-detection experiments such as the upcoming 
XENON1-T~\cite{Aprile:2012zx} and superCDMS~\cite{Akerib:2006rr} 
experiments which will probe the spin-indepedent cross-sections for WIMP dark matter.  
The spin-independent direct-detection cross-sections are plotted vs. higgsino fraction for the neutralino LSP
in the lower panel of Fig.~\ref{fig:DMdensityvsHiggsinoFractiontb30mupos}.  Here, it can be clearly seen that the cross-sections
become much smaller with a larger higgsino fraction.  This can be easily understood as being due
to the weakness with which higgsinos couple.  
In Fig.~\ref{fig:DirectDetectCrossSectionstb30mupos}, the re-scaled spin-independent cross-sections are
plotted vs. LSP mass, where the cross-sections have been re-scaled by the ratio of the neutralino relic density to 
the upper limit on the 9-year $2\sigma$ WMAP bound, $\Omega_{DM} h^2 <= 0.123$.  This has the effect of lowering
the effective cross-section on the pure higgsino LSP by roughly a factor of ten since the relic density 
in this case is roughly a factor of ten lower than the dark matter density observed by WMAP and Planck. Also shown
on the Figure is the XENON100 upper limit on the SI proton-neutralino 
cross-section~\cite{Aprile:2012nq,Aprile:2011hi} as well as the projected reach of the next generation
XENON1-T experiment~\cite{Aprile:2012zx}.   
In the Figure, it can be seen that spectra with a higher LSP mass are ruled out or nearly ruled out by the most
recent XENON100 results.  Other points in the parameter space should be completely covered by the upcoming 
XENON1-T experiment including those with a light higgsino LSP, which are just within reach.  
However, the spin-dependent direct-detection cross-sections may not be in reach as shown in 
Fig.~\ref{fig:SDDirectDetectCrossSectionstb30mupos}~\cite{Garny:2012it,Baudis:2012bc}.

Also of great interest are the prospects for indirect dark matter detection resulting from
neutralino dark matter annihilations.  The annihilation cross sections to continuum photons 
are shown in Fig.~\ref{fig:AnnihilationCrossSections} for model points satisfying $\Omega_{DM} h^2 <= 0.123$
with $\Delta m^2 = 0$, tan$\beta = 30$, $\mu > 0$, and $m_t=173$~GeV.  For comparison, the constraints
derived from a combined analysis of dwarf spheroidal (dSph) Milky Way 
galaxies~\cite{Ackermann:2011wa,GeringerSameth:2011iw} are also shown.
As can be seen from this plot most of the model points, specifically those 
with a neutralino mass $m_{\tilde{\chi}_0} \gtrsim 250$~GeV,
are below the Fermi dSph constraint.

\section{Conclusion}

When supersymmetry breaking is dominated by 
the complex structure moduli and the universal dilaton,
a subset of the supersymmetry parameter space in a semi-realistic MSSM constructed from 
intersecting/magnetized D-branes is 
similar to the mSUGRA/CMSSM parameter space with the trilinear term fixed to be 
minus the gaugino mass, $A_0=-m_{1/2}$.   
More generally, the scalar mass-squared
terms for sfermions are split about the Higgs mass-squared terms,
$m_{Q_L,L_L}^2=m_H^2 - \Delta m^2$ and $m_{Q_R,L_R}^2=m_H^2 + \Delta m^2$,
for generic values of the K\a"ahler moduli.  The hyberbolic branch/focus point (HB/FP)
regions of this parameter space are present for both $\Delta m^2 = 0$ and $\Delta m^2 \ne 0$. 
In this work, we have studied these regions in detail.  

In the case of the high-scale boundary conditions obtained from the model 
for supersupersymmetry breaking dominated by the complex structure moduli
and the dilaton, it is very interesting that they take the form 
$(m^2_{H_u}, m^2_{U_3}, m^2_{Q_3})\propto(1,1+x,1-x)$ with $x$ an
arbitrary constant.  In the present context, x is a function of the 
K\a"ahler moduli in context of this model.  As was discussed, it has
been known for some time that focus points may be obtained for any 
boundary conditions of this form.  Thus, the HB/FP region discussed 
for $\Delta m^2 = x = 0$ remains present even when $\Delta m^2 = x \ne 0$.  
Our results verified that this was indeed the case.  With respect to the 
intersecting/magnetized D-brane model, this implies that the resulting physics 
does not depend upon the choice of K\a"ahler moduli.  

It has been shown that 
there exists superpartner spectra with a light Higgsino-like LSP with a mass in the range 
$230-350$~GeV and a Higgs
mass in the range $124-126$~GeV, and which satisfy most standard experimental constraints.  
Consequently, viable spectra
with low EWFT between $3-7\%$ may be obtained.  The rescaled spin-independent direct-detection
crosssections are in range of future experiments such as XENON-1T and super CDMS, while
the relic density is smaller than the WMAP and Planck bounds by roughly a factor of ten. 
For superpartner spectra which fall within the WMAP 9-year $2\sigma$ bounds, the LSP mass
is in the range $450-750$~GeV. These spectra require a minimum fine-tuning of $1-2\%$.  
In addition, a portion of this parameter space is already excluded by the 
XENON100 results.  Thus, the parameter space with a nearly pure higgsino composition
is favored.  However this implies that the LSP provides $\sim10\%$ of the
dark matter density observed by WMAP and Planck. Axions or
superheavy hidden-sector dark matter may then provide the bulk of the astrophysical dark matter.
Also of great interest are the prospects for indirect dark matter detection resulting from
neutralino dark matter annihilations.  The annihilation cross sections to continuum photons 
for most of the model points, specifically those 
with a neutralino mass $m_{\tilde{\chi}_0} \gtrsim 250$~GeV,
are below the Fermi dSph constraint.

For all of these spectra, the gluino mass and the scalar masses are all in the multi-TeV
range.  In particular, the gluino mass is in the range $2.6-4.4$~TeV while the lightest
stop mass is in the range $3.5-5.6$~TeV.  The other squark and slepton masses are typically
in the $5-10$~TeV range, as are the masses of the additional Higgs scalars.  
Due to the heavy masses for the gluino and squarks in these models, it would be
very difficult to observe a clear and distinct signal at the LHC 
if the spectrum of superpartners falls into these regions of the
parameter space.  However, the prospects for observing superpartners 
at a linear collider or at a higher-energy hadron collider 
appear to be more promising. We shall soon see as LHC run II
is currently underway at a higher CM energy of $13$~TeV.

\section{Acknowledgments}
VEM would like to thank the University of Houston-Clear Lake Natural Sciences division for the purchase of a high-performance
Linux workstation which was used to complete this work.


\newpage

\begin{thebibliography}{99}

\itemsep 0.5mm

 

\bibitem{Buchmuller:2005jr}
  W.~Buchmuller, K.~Hamaguchi, O.~Lebedev and M.~Ratz,
  Phys.\ Rev.\ Lett.\  {\bf 96}, 121602 (2006),
  and references therein.


\bibitem{Lebedev:2006kn}
  O.~Lebedev, H.~P.~Nilles, S.~Raby, S.~Ramos-Sanchez, M.~Ratz, 
P.~K.~S.~Vaudrevange and A.~Wingerter,
  Phys.\ Lett.\  B {\bf 645}, 88 (2007), and references therein.


\bibitem{Kim:2006hw}
  J.~E.~Kim and B.~Kyae,
  Nucl.\ Phys.\  B {\bf 770}, 47 (2007);
  Phys.\ Rev.\  D {\bf 77}, 106008 (2008);
J.~H.~Huh, J.~E.~Kim and B.~Kyae,
  arXiv:0904.1108 [hep-ph].



\bibitem{Braun:2005ux}
  V.~Braun, Y.~H.~He, B.~A.~Ovrut and T.~Pantev,
  Phys.\ Lett.\  B {\bf 618}, 252 (2005);
  JHEP {\bf 0605}, 043 (2006),  and references therein.


\bibitem{Bouchard:2005ag}
  V.~Bouchard and R.~Donagi,
  Phys.\ Lett.\  B {\bf 633}, 783 (2006), 
  and references therein.


\bibitem{AEHN}
I.~Antoniadis, J.~R.~Ellis, J.~S.~Hagelin and D.~V.~Nanopoulos,
Phys.\ Lett.\ B {\bf 205} (1988) 459;
Phys.\ Lett.\ B {\bf 208} (1988) 209 [Addendum-ibid.\ B {\bf 213}
(1988) 562];
Phys.\ Lett.\ B {\bf 231} (1989) 65.


\bibitem{Faraggi:1989ka}
  A.~E.~Faraggi, D.~V.~Nanopoulos and K.~J.~Yuan,
  Nucl.\ Phys.\ B {\bf 335}, 347 (1990).


\bibitem{Antoniadis:1990hb}
  I.~Antoniadis, G.~K.~Leontaris and J.~Rizos,
  Phys.\ Lett.\ B {\bf 245}, 161 (1990).


\bibitem{LNY}
  J.~L.~Lopez, D.~V.~Nanopoulos and K.~J.~Yuan,
  Nucl.\ Phys.\ B {\bf 399}, 654 (1993);
  D.~V.~Nanopoulos,
  hep-ph/0211128.


\bibitem{Cleaver:2001ab}
  G.~B.~Cleaver, A.~E.~Faraggi, D.~V.~Nanopoulos and J.~W.~Walker,
  Nucl.\ Phys.\  B {\bf 620}, 259 (2002), and references therein.


\bibitem{Berkooz:1996km}
  M.~Berkooz, M.~R.~Douglas and R.~G.~Leigh,
  Nucl.\ Phys.\  B {\bf 480}, 265 (1996).

\bibitem{Ibanez:2001nd}
  L.~E.~Ibanez, F.~Marchesano and R.~Rabadan,
  JHEP {\bf 0111}, 002 (2001).

\bibitem{Blumenhagen:2001te}
  R.~Blumenhagen, B.~Kors, D.~Lust and T.~Ott,
  Nucl.\ Phys.\  B {\bf 616}, 3 (2001).

\bibitem{CSU}
M.~Cveti\v c, G.~Shiu and A.~M.~Uranga, Phys.\ Rev.\ Lett.\  {\bf
87}, 201801 (2001);
M.~Cveti\v c, G.~Shiu and A.~M.~Uranga, Nucl.\ Phys.\ B {\bf 615},
3 (2001).


\bibitem{Cvetic:2002pj}
  M.~Cveti\v c, I.~Papadimitriou and G.~Shiu,
  Nucl.\ Phys.\ B {\bf 659}, 193 (2003)
  [Erratum-ibid.\ B {\bf 696}, 298 (2004)].

\bibitem{Cvetic:2004ui} 
  M.~Cvetic, T.~Li and T.~Liu,
  Nucl.\ Phys.\ B {\bf 698}, 163 (2004)
  [hep-th/0403061].


\bibitem{Cvetic:2004nk} 
  M.~Cvetic, P.~Langacker, T.~Li and T.~Liu,
  Nucl.\ Phys.\ B {\bf 709}, 241 (2005)
  [hep-th/0407178].


\bibitem{Cvetic:2005bn} 
  M.~Cvetic, T.~Li and T.~Liu,
  Phys.\ Rev.\ D {\bf 71}, 106008 (2005)
  [hep-th/0501041].


\bibitem{Chen:2005ab}
  C.-M.~Chen, G.~V.~Kraniotis, V.~E.~Mayes, D.~V.~Nanopoulos and J.~W.~Walker,
  Phys.\ Lett.\ B {\bf 611}, 156 (2005);
  Phys.\ Lett.\  B {\bf 625}, 96 (2005).


\bibitem{Chen:2005mj}
  C.~M.~Chen, T.~Li and D.~V.~Nanopoulos,
  Nucl.\ Phys.\ B {\bf 732}, 224 (2006).


\bibitem{Blumenhagen:2005mu}
  R.~Blumenhagen, M.~Cvetic, P.~Langacker and G.~Shiu,
  Ann.\ Rev.\ Nucl.\ Part.\ Sci.\  {\bf 55}, 71 (2005),
and references therein.


\bibitem{Dijkstra:2004ym}
  T.~P.~T.~Dijkstra, L.~R.~Huiszoon and A.~N.~Schellekens,
  Phys.\ Lett.\  B {\bf 609}, 408 (2005).


\bibitem{Dijkstra:2004cc}
  T.~P.~T.~Dijkstra, L.~R.~Huiszoon and A.~N.~Schellekens,
  Nucl.\ Phys.\  B {\bf 710}, 3 (2005), and references therein.

\bibitem{Mayes:2013bda} 
  V.~E.~Mayes,
  Nucl.\ Phys.\ B {\bf 877}, 401 (2013)
  [arXiv:1305.2842 [hep-ph]].
	
\bibitem{Chen:2007zu} 
  C.~-M.~Chen, T.~Li, V.~E.~Mayes and D.~V.~Nanopoulos,
  Phys.\ Rev.\ D {\bf 77}, 125023 (2008)
  [arXiv:0711.0396 [hep-ph]].
	
	\bibitem{Chen:2007px} 
  C.~-M.~Chen, T.~Li, V.~E.~Mayes and D.~V.~Nanopoulos,
  Phys.\ Lett.\ B {\bf 665}, 267 (2008)
  [hep-th/0703280].


\bibitem{Carena:2002es} 
  M.~S.~Carena and H.~E.~Haber,
  Prog.\ Part.\ Nucl.\ Phys.\  {\bf 50}, 63 (2003)
  [hep-ph/0208209].

\bibitem{Ellis:1983wd} 
  J.~R.~Ellis, J.~S.~Hagelin, D.~V.~Nanopoulos and M.~Srednicki,
  Phys.\ Lett.\ B {\bf 127}, 233 (1983).

\bibitem{Ellis:1983ew} 
  J.~R.~Ellis, J.~S.~Hagelin, D.~V.~Nanopoulos, K.~A.~Olive and M.~Srednicki,
  Nucl.\ Phys.\ B {\bf 238}, 453 (1984).

\bibitem{Ellis:1982wr}
  J.~R.~Ellis, D.~V.~Nanopoulos and K.~Tamvakis,
  Phys.\ Lett.\  B {\bf 121}, 123 (1983).
  
 \bibitem{Dimopoulos:1981yj} 
  S.~Dimopoulos, S.~Raby and F.~Wilczek,
  Phys.\ Rev.\ D {\bf 24}, 1681 (1981).
  
  \bibitem{Ibanez:1981yh} 
  L.~E.~Ibanez and G.~G.~Ross,
  Phys.\ Lett.\ B {\bf 105}, 439 (1981).
  
	\bibitem{:2012gk} 
  G.~Aad {\it et al.}  [ATLAS Collaboration],
  Phys.\ Lett.\ B {\bf 716}, 1 (2012)
  [arXiv:1207.7214 [hep-ex]].

\bibitem{:2012gu} 
  S.~Chatrchyan {\it et al.}  [CMS Collaboration],
  Phys.\ Lett.\ B {\bf 716}, 30 (2012)
  [arXiv:1207.7235 [hep-ex]].

	
  \bibitem{Aad:2012fqa} 
  G.~Aad {\it et al.}  [ATLAS Collaboration],
  Phys.\ Rev.\ D {\bf 87}, 012008 (2013)
  [arXiv:1208.0949 [hep-ex]].
 
 \bibitem{Aad:2012hm} 
  G.~Aad {\it et al.}  [ATLAS Collaboration],
  JHEP {\bf 1207}, 167 (2012)
  [arXiv:1206.1760 [hep-ex]].
 
 \bibitem{:2012mfa} 
  S.~Chatrchyan {\it et al.}  [CMS Collaboration],
  Phys.\ Rev.\ Lett.\  {\bf 109}, 171803 (2012)
  [arXiv:1207.1898 [hep-ex]].
  
 \bibitem{Aad:2011ib} 
  G.~Aad {\it et al.}  [ATLAS Collaboration],
  Phys.\ Lett.\ B {\bf 710}, 67 (2012)
  [arXiv:1109.6572 [hep-ex]].
  
 \bibitem{Chatrchyan:2011zy} 
  S.~Chatrchyan {\it et al.}  [CMS Collaboration],
  Phys.\ Rev.\ Lett.\  {\bf 107}, 221804 (2011)
  [arXiv:1109.2352 [hep-ex]].
 
\bibitem{Chamseddine:1982jx} 
  A.~H.~Chamseddine, R.~L.~Arnowitt and P.~Nath,
  Phys.\ Rev.\ Lett.\  {\bf 49}, 970 (1982).

\bibitem{Ohta:1982wn} 
  N.~Ohta,
  Prog.\ Theor.\ Phys.\  {\bf 70}, 542 (1983).  
  
\bibitem{Hall:1983iz} 
  L.~J.~Hall, J.~D.~Lykken and S.~Weinberg,
  Phys.\ Rev.\ D {\bf 27}, 2359 (1983).

\bibitem{Chan:1997bi} 
  K.~L.~Chan, U.~Chattopadhyay and P.~Nath,
  Phys.\ Rev.\ D {\bf 58}, 096004 (1998)
  [hep-ph/9710473].
  
 
\bibitem{Feng:1999mn} 
  J.~L.~Feng, K.~T.~Matchev and T.~Moroi,
  Phys.\ Rev.\ Lett.\  {\bf 84}, 2322 (2000)
  [hep-ph/9908309].

\bibitem{Feng:1999zg} 
  J.~L.~Feng, K.~T.~Matchev and T.~Moroi,
  Phys.\ Rev.\ D {\bf 61}, 075005 (2000)
  [hep-ph/9909334].
  
\bibitem{Baer:1995nq} 
  H.~Baer, C.~-h.~Chen, F.~Paige and X.~Tata,
  Phys.\ Rev.\ D {\bf 52}, 2746 (1995)
  [hep-ph/9503271].

\bibitem{Baer:1998sz} 
  H.~Baer, C.~-h.~Chen, M.~Drees, F.~Paige and X.~Tata,
  Phys.\ Rev.\ D {\bf 59}, 055014 (1999)
  [hep-ph/9809223].

\bibitem{Chattopadhyay:2003xi} 
  U.~Chattopadhyay, A.~Corsetti and P.~Nath,
  Phys.\ Rev.\ D {\bf 68}, 035005 (2003)
  [hep-ph/0303201]. 

\bibitem{Kadastik:2011aa} 
  M.~Kadastik, K.~Kannike, A.~Racioppi and M.~Raidal,
  JHEP {\bf 1205}, 061 (2012)
  [arXiv:1112.3647 [hep-ph]].

\bibitem{Strege:2011pk} 
  C.~Strege, G.~Bertone, D.~G.~Cerdeno, M.~Fornasa, R.~Ruiz de Austri and R.~Trotta,
  JCAP {\bf 1203}, 030 (2012)
  [arXiv:1112.4192 [hep-ph]].

\bibitem{Aparicio:2012iw} 
  L.~Aparicio, D.~G.~Cerdeno and L.~E.~Ibanez,
  JHEP {\bf 1204}, 126 (2012)
  [arXiv:1202.0822 [hep-ph]].
  
\bibitem{Ellis:2012aa} 
  J.~Ellis and K.~A.~Olive,
  Eur.\ Phys.\ J.\ C {\bf 72}, 2005 (2012)
  [arXiv:1202.3262 [hep-ph]]. 

\bibitem{Baer:2012uya} 
  H.~Baer, V.~Barger and A.~Mustafayev,
  JHEP {\bf 1205}, 091 (2012)
  [arXiv:1202.4038 [hep-ph]].
  
\bibitem{Matchev:2012vf} 
  K.~Matchev and R.~Remington,
  arXiv:1202.6580 [hep-ph].

\bibitem{Akula:2012kk} 
  S.~Akula, P.~Nath and G.~Peim,
  Phys.\ Lett.\ B {\bf 717}, 188 (2012)
  [arXiv:1207.1839 [hep-ph]].

\bibitem{Ghosh:2012dh} 
  D.~Ghosh, M.~Guchait, S.~Raychaudhuri and D.~Sengupta,
  Phys.\ Rev.\ D {\bf 86}, 055007 (2012)
  [arXiv:1205.2283 [hep-ph]].
  
\bibitem{Fowlie:2012im} 
  A.~Fowlie, M.~Kazana, K.~Kowalska, S.~Munir, L.~Roszkowski, E.~M.~Sessolo, S.~Trojanowski and Y.~-L.~S.~Tsai,
  Phys.\ Rev.\ D {\bf 86}, 075010 (2012)
  [arXiv:1206.0264 [hep-ph]].

\bibitem{Buchmueller:2012hv} 
  O.~Buchmueller, R.~Cavanaugh, M.~Citron, A.~De Roeck, M.~J.~Dolan, J.~R.~Ellis, H.~Flacher and S.~Heinemeyer {\it et al.},
  Eur.\ Phys.\ J.\ C {\bf 72}, 2243 (2012)
  [arXiv:1207.7315 [hep-ph]].
  
\bibitem{Strege:2012bt} 
  C.~Strege, G.~Bertone, F.~Feroz, M.~Fornasa, R.~Ruiz de Austri and R.~Trotta,
  JCAP {\bf 1304}, 013 (2013)
  [arXiv:1212.2636 [hep-ph]].
  
\bibitem{Citron:2012fg} 
   M.~Citron, J.~Ellis, F.~Luo, J.~Marrouche, K.~A.~Olive and K.~J.~de Vries,
  Phys.\ Rev.\ D {\bf 87}, 036012 (2013)
  [arXiv:1212.2886 [hep-ph]].
  
\bibitem{Ellis:2012nv} 
  J.~Ellis, F.~Luo, K.~A.~Olive and P.~Sandick,
  Eur.\ Phys.\ J.\ C {\bf 73}, no. 4, 2403 (2013)
  [arXiv:1212.4476 [hep-ph]].

\bibitem{Boehm:2012rh} 
  C.~Boehm, J.~Da Silva, A.~Mazumdar and E.~Pukartas,
  Phys.\ Rev.\ D {\bf 87}, 023529 (2013)  
  [arXiv:1205.2815 [hep-ph]].


\bibitem{Baer:2012mv} 
  H.~Baer, V.~Barger, P.~Huang, D.~Mickelson, A.~Mustafayev and X.~Tata,
  Phys.\ Rev.\ D {\bf 87}, no. 3, 035017 (2013)
  [arXiv:1210.3019 [hep-ph]].

\bibitem{Ellis:2002wv} 
  J.~R.~Ellis, K.~A.~Olive and Y.~Santoso,
  Phys.\ Lett.\ B {\bf 539}, 107 (2002)
  [hep-ph/0204192].

\bibitem{Ellis:2002iu} 
  J.~R.~Ellis, T.~Falk, K.~A.~Olive and Y.~Santoso,
  Nucl.\ Phys.\ B {\bf 652}, 259 (2003)
  [hep-ph/0210205].

\bibitem{Baer:2004fu} 
  H.~Baer, A.~Mustafayev, S.~Profumo, A.~Belyaev and X.~Tata,
  Phys.\ Rev.\ D {\bf 71}, 095008 (2005)
  [hep-ph/0412059].

\bibitem{Baer:2005bu} 
  H.~Baer, A.~Mustafayev, S.~Profumo, A.~Belyaev and X.~Tata,
  JHEP {\bf 0507}, 065 (2005)
  [hep-ph/0504001].

\bibitem{Baer:2013xua} 
  H.~Baer, V.~Barger, P.~Huang, D.~Mickelson, A.~Mustafayev, W.~Sreethawong and X.~Tata,
  JHEP {\bf 1312}, 013 (2013)
  [arXiv:1310.4858 [hep-ph]]

\bibitem{Maxin:2009ez} 
  J.~A.~Maxin, V.~E.~Mayes and D.~V.~Nanopoulos,
  Phys.\ Rev.\ D {\bf 81}, 015008 (2010)
  [arXiv:0908.0915 [hep-ph]].
	
\bibitem{Li:2014xqa} 
  T.~Li, D.~V.~Nanopoulos, S.~Raza and X.~C.~Wang,
  JHEP {\bf 1408}, 128 (2014)
  [arXiv:1406.5574 [hep-ph]].


\bibitem{Aprile:2012zx} 
  E.~Aprile [XENON1T Collaboration],
  Springer Proc.\ Phys.\  {\bf 148}, 93 (2013)
  [arXiv:1206.6288 [astro-ph.IM]].

\bibitem{Akerib:2006rr} 
  D.~S.~Akerib, M.~J.~Attisha, C.~N.~Bailey, L.~Baudis, D.~A.~Bauer, P.~L.~Brink, P.~P.~Brusov and R.~Bunker {\it et al.},
  Nucl.\ Instrum.\ Meth.\ A {\bf 559}, 411 (2006).

\bibitem{Maxin:2011ne} 
  J.~A.~Maxin, V.~E.~Mayes and D.~V.~Nanopoulos,
  Phys.\ Rev.\ D {\bf 84}, 106009 (2011)
  [arXiv:1108.0887 [hep-ph]].

\bibitem{Mayes:2013qmc} 
  V.~E.~Mayes,
  Int.\ J.\ Mod.\ Phys.\ A {\bf 28}, 1350061 (2013)
  [arXiv:1302.4394 [hep-ph]].
	
\bibitem{Drees:2004jm} 
  M.~Drees, R.~Godbole and P.~Roy,
  Hackensack, USA: World Scientific (2004) 555 p
	

\bibitem{Blumenhagen:2006ci}
  R.~Blumenhagen, B.~Kors, D.~Lust and S.~Stieberger,
  Phys.\ Rept.\  {\bf 445}, 1 (2007)
  [arXiv:hep-th/0610327].
	

\bibitem{Cremades:2002te}
  D.~Cremades, L.~E.~Ib\'{a}\~{n}ez and F.~Marchesano,
  JHEP {\bf 0207}, 009 (2002)
  [arXiv:hep-th/0201205].


\bibitem{Shiu:1998pa}
  G.~Shiu and S.~H.~H.~Tye,
  Phys.\ Rev.\  D {\bf 58}, 106007 (1998)
  [arXiv:hep-th/9805157].

\bibitem{CLS1}
 M.~Cveti\v c, P.~Langacker and G.~Shiu,
Phys.\ Rev.\ D {\bf 66}, 066004 (2002);
 Nucl.\ Phys.\ B {\bf 642}, 139 (2002).

\bibitem{Lust:2003ky}
  D.~L\"ust and S.~Stieberger,
  [arXiv:hep-th/0302221].


\bibitem{Antoniadis:Blumen}
  I.~Antoniadis, E.~Kiritsis and T.~N.~Tomaras,
  Phys.\ Lett.\  B {\bf 486}, 186 (2000)
  [arXiv:hep-ph/0004214];
  R.~Blumenhagen, D.~Lust and S.~Stieberger,
  JHEP {\bf 0307}, 036 (2003)
  [arXiv:hep-th/0305146].

\bibitem{Cremades:2003qj}
  D.~Cremades, L.~E.~Ib\'{a}\~{n}ez and F.~Marchesano,
  JHEP {\bf 0307}, 038 (2003)
  [arXiv:hep-th/0302105].


\bibitem{Cvetic:2003ch}
  M.~Cveti\v{c} and I.~Papadimitriou,
  Phys.\ Rev.\  D {\bf 68}, 046001 (2003)
  [Erratum-ibid.\  D {\bf 70}, 029903 (2004)]
  [arXiv:hep-th/0303083].

\bibitem{Kors:2003wf}
  B.~K\"ors and P.~Nath,
  Nucl.\ Phys.\  B {\bf 681}, 77 (2004)
  [arXiv:hep-th/0309167].

\bibitem{Lust:2004cx}

  D.~L\"ust, P.~Mayr, R.~Richter and S.~Stieberger,
  Nucl.\ Phys.\  B {\bf 696}, 205 (2004);
 A.~Font and L.\,E.~Ibanez,
  JHEP {\bf 0503}, 040 (2005).
	
	\bibitem{Kawamura:1996ex}
  Y.~Kawamura, T.~Kobayashi and T.~Komatsu,
  Phys.\ Lett.\  B {\bf 400}, 284 (1997)
  [arXiv:hep-ph/9609462].

	
  \bibitem{Maxin:2008kp} 
  J.~A.~Maxin, V.~E.~Mayes and D.~V.~Nanopoulos,
  Phys.\ Rev.\ D {\bf 79}, 066010 (2009)
  [arXiv:0809.3200 [hep-ph]].
		
\bibitem{Maxin:2009pr} 
  J.~A.~Maxin, V.~E.~Mayes and D.~V.~Nanopoulos,
  Phys.\ Lett.\ B {\bf 690}, 501 (2010)
  [arXiv:0911.2806 [hep-ph]].

\bibitem{Ellis:1983bp}
  J.~R.~Ellis, J.~S.~Hagelin, D.~V.~Nanopoulos and K.~Tamvakis,
  Phys.\ Lett.\  B {\bf 125}, 275 (1983).

\bibitem{AlvarezGaume:1983gj}
  L.~Alvarez-Gaume, J.~Polchinski and M.~B.~Wise,
  Nucl.\ Phys.\  B {\bf 221}, 495 (1983).

\bibitem{SUSP}
   A.~Djouadi, J.~Kneur, and G.~Moultaka,
  Comput.\ Phys.\ Commun. {\bf 176}, (2007) 426-455.
  arXiv:hep-ph/0211331v2
	 
  \bibitem{Belanger:2010pz}
  G.~Belanger, F.~Boudjema, A.~Pukhov and A.~Semenov,
  arXiv:1005.4133 [hep-ph].
  
  \bibitem{Belanger:2008sj}
  G.~Belanger, F.~Boudjema, A.~Pukhov and A.~Semenov,
  arXiv:0803.2360 [hep-ph].

  \bibitem{Belanger:2006is}
  G.~Belanger, F.~Boudjema, A.~Pukhov and A.~Semenov,
  Comput.\ Phys.\ Commun.\  {\bf 176} (2007) 367
  [arXiv:hep-ph/0607059].

  
\bibitem{ATLAS:2014wva} 
  [ATLAS and CDF and CMS and D0 Collaborations],
  arXiv:1403.4427 [hep-ex].
 
\bibitem{Ade:2013lta}  P.~A.~R.~Ade {\it et al.}  [Planck Collaboration],
  Astron.\ Astrophys.\  {\bf 571}, A16 (2014)
  [arXiv:1303.5076 [astro-ph.CO]].
  P.~A.~R.~Ade {\it et al.}  [Planck Collaboration],
  arXiv:1303.5076 [astro-ph.CO]. 

\bibitem{Hinshaw:2012fq} 
   G.~Hinshaw {\it et al.}  [WMAP Collaboration],
  Astrophys.\ J.\ Suppl.\  {\bf 208}, 19 (2013)
  [arXiv:1212.5226 [astro-ph.CO]].
 
  
\bibitem{HFAG}
  E.~Barberio, \textit{et al} (Heavy Flavor Averaging Group),
  arXiv:hep-ex/0704.3575v1

\bibitem{MMEA}
  M.~Misiak \textit{et al},
  Phys.\ Rev.\  Lett. {\bf 98}, 022002 (2007).
  arXiv:hep-ph/0609232v2
 
\bibitem{:2012ct} 
  RAaij {\it et al.}  [LHCb Collaboration],
  Phys.\ Rev.\ Lett.\  {\bf 110}, 021801 (2013)
  [arXiv:1211.2674].

\bibitem{MUON}
  G.~W.~Bennett \textit{et al} (Muon g-2 Collaboration),
  Phys.\ Rev.\ Lett. {\bf 92}, 161802 (2004).
  arXiv:hep-ex/0401008

\bibitem{Davier:2010nc} 
  M.~Davier, A.~Hoecker, B.~Malaescu and Z.~Zhang,
  Eur.\ Phys.\ J.\ C {\bf 71}, 1515 (2011)
  [Erratum-ibid.\ C {\bf 72}, 1874 (2012)]
  [arXiv:1010.4180 [hep-ph]].

\bibitem{Bae:2013bva} 
  K.~J.~Bae, H.~Baer and E.~J.~Chun,
  Phys.\ Rev.\ D {\bf 89}, no. 3, 031701 (2014)
  [arXiv:1309.0519 [hep-ph]].
	
\bibitem{Ellis:1990iu} 
  J.~R.~Ellis, J.~L.~Lopez and D.~V.~Nanopoulos,
  Phys.\ Lett.\ B {\bf 247}, 257 (1990).	

\bibitem{Kolb}
E.~W.~Kolb, A.~D.~Linde and A.~Riotto,
Phys.\ Rev.\ Lett.\  {\bf 77} (1996) 4290
[arXiv:hep-ph/9606260];
B.~R.~Greene, T.~Prokopec and T.~G.~Roos,
Phys.\ Rev.\ D {\bf 56} (1997) 6484
[arXiv:hep-ph/9705357].
E.~W.~Kolb, A.~Riotto and I.~I.~Tkachev,
Phys.\ Lett.\ B {\bf 423} (1998) 348
[arXiv:hep-ph/9801306];
D.~J.~H.~Chung, E.~W.~Kolb and A.~Riotto,
Phys.\ Rev.\ D {\bf 59} (1999) 023501
[arXiv:hep-ph/9802238].


\bibitem{Baer:2011ec} 
  H.~Baer, V.~Barger and P.~Huang,
  JHEP {\bf 1111}, 031 (2011)
  [arXiv:1107.5581 [hep-ph]].


	
\bibitem{Aprile:2011hi} 
  E.~Aprile {\it et al.}  [XENON100 Collaboration],
  Phys.\ Rev.\ Lett.\  {\bf 107}, 131302 (2011)
  [arXiv:1104.2549 [astro-ph.CO]].

\bibitem{Aprile:2012nq} 
  E.~Aprile {\it et al.}  [XENON100 Collaboration],
  Phys.\ Rev.\ Lett.\  {\bf 109}, 181301 (2012)
  [arXiv:1207.5988 [astro-ph.CO]].
	
\bibitem{Garny:2012it} 
  M.~Garny, A.~Ibarra, M.~Pato and S.~Vogl,
  Phys.\ Rev.\ D {\bf 87}, no. 5, 056002 (2013)
  [arXiv:1211.4573 [hep-ph]].

\bibitem{Baudis:2012bc} 
  L.~Baudis [DARWIN Consortium Collaboration],
  J.\ Phys.\ Conf.\ Ser.\  {\bf 375}, 012028 (2012)
  [arXiv:1201.2402 [astro-ph.IM]].	


\bibitem{Ackermann:2011wa} 
  M.~Ackermann {\it et al.}  [Fermi-LAT Collaboration],
  Phys.\ Rev.\ Lett.\  {\bf 107}, 241302 (2011)
  [arXiv:1108.3546 [astro-ph.HE]].
	
\bibitem{GeringerSameth:2011iw} 
  A.~Geringer-Sameth and S.~M.~Koushiappas,
  Phys.\ Rev.\ Lett.\  {\bf 107}, 241303 (2011)
  [arXiv:1108.2914 [astro-ph.CO]].


\end{thebibliography}
\end{document}